\pdfoutput=1
\documentclass[sigconf]{acmart} 

\AtBeginDocument{%
  \providecommand\BibTeX{{%
    \normalfont B\kern-0.5em{\scshape i\kern-0.25em b}\kern-0.8em\TeX}}}

\copyrightyear{2021}
\acmYear{2021}
\setcopyright{acmlicensed}
\acmConference[SIGIR '21]{Proceedings of the 44th International ACM SIGIR Conference on Research and Development in Information Retrieval}{July 11--15, 2021}{Virtual Event, Canada} 
\acmBooktitle{Proceedings of the 44th International ACM SIGIR Conference on Research and Development in Information Retrieval (SIGIR '21), July 11--15, 2021, Virtual Event, Canada}
\acmPrice{15.00}
\acmDOI{10.1145/3404835.3462886}
\acmISBN{978-1-4503-8037-9/21/07}

\usepackage{amsmath,amsfonts}
\usepackage{algorithm}
\usepackage{algorithmic}
\usepackage{bm}
\usepackage{color}
\usepackage{comment}
\usepackage{graphicx}
\usepackage{multirow}
\usepackage{subfigure}
\usepackage{siunitx}
\usepackage{pifont}

\begin{document}
\fancyhead{}

\title{Set2setRank: Collaborative Set to Set Ranking  for Implicit Feedback based Recommendation}

% 1Key Laboratory of Knowledge Engineering with Big Data, Hefei University of Technology 
% 2 School of Computer Science and Information Engineering, Hefei University of Technology
% 3. Institute of Artificial Intelligence, Hefei Comprehensive National Science Center

% \settopmatter{authorsperrow=5}

\author{Lei Chen$^{1,2}$, Le Wu$^{1,2,3,*}$, Kun Zhang$^{1,2}$, Richang Hong$^{1,2}$, Meng Wang$^{1,2,3}$}
\affiliation[obeypunctuation=true]{\institution{$^1$ Key Laboratory of Knowledge Engineering with Big Data, Hefei University of Technology,\country{China}}}
\affiliation[obeypunctuation=true]{\institution{$^2$ School of Computer Science and Information Engineering, Hefei University of Technology,\country{China}}}
\affiliation[obeypunctuation=true]{\institution{$^3$ Institute of Artificial Intelligence, Hefei Comprehensive National Science Center,\country{China}}}
\email{{chenlei.hfut, lewu.ustc, zhang1028kun, hongrc.hfut, eric.mengwang}@gmail.com}

\thanks{Le Wu is the corresponding author.}

\begin{abstract}
As users often express their preferences with binary behavior data~(implicit feedback), such as clicking items or buying products, implicit feedback based Collaborative Filtering~(CF) models predict the top ranked items a user might like by leveraging implicit user-item interaction data. For each user, the implicit feedback is divided into two sets: an observed item set with limited observed behaviors, and a large unobserved item set that is mixed with negative item behaviors and unknown behaviors. Given any user preference prediction model, researchers either designed ranking based optimization goals or relied on negative item mining techniques for better optimization. Despite the performance gain of these implicit feedback based models, the recommendation results are still far from satisfactory due to the sparsity of the observed item set for each user. To this end, in this paper, we explore the unique characteristics of the implicit feedback and propose Set2setRank framework for recommendation. The optimization criteria of Set2setRank are two folds: First, we design an item to an item set comparison that encourages each observed item from the sampled observed set is ranked higher than any unobserved item from the sampled unobserved set. Second, we model set level comparison that encourages a margin between the distance summarized from the observed item set and the most ``hard'' unobserved item from the sampled negative set. Further, an adaptive sampling technique is designed to implement these two goals. We have to note that our proposed framework is model-agnostic and can be easily applied to most recommendation prediction approaches, and is time efficient in practice. Finally, extensive experiments on three real-world datasets demonstrate the superiority of our proposed approach. 

\end{abstract}

\begin{CCSXML}
<ccs2012>
   <concept>
       <concept_id>10002951.10003317.10003347.10003350</concept_id>
       <concept_desc>Information systems~Recommender systems</concept_desc>
       <concept_significance>500</concept_significance>
       </concept>
   <concept>
       <concept_id>10002951.10003317.10003338.10003343</concept_id>
       <concept_desc>Information systems~Learning to rank</concept_desc>
       <concept_significance>500</concept_significance>
       </concept>
 </ccs2012>
\end{CCSXML}

\ccsdesc[500]{Information systems~Recommender systems}
\ccsdesc[500]{Information systems~Learning to rank}

\keywords{recommendation; implicit feedback; collaborative ranking; adaptive sampling; self-supervised ranking}

\maketitle

\section{Introduction}

Collaborative Filtering~(CF) provides personalized ranking list for each user by leveraging the collaborative signals from user-item interaction data, and are popular in most recommender systems with easy to collect data and relatively high performance~\cite{wu2019noise,wang2019unified,chae2020ar,wu2020joint}. Earlier approaches worked on  the explicit user feedback data~(e.g., 1-5 rating scale), and relied on pointwise optimization functions by comparing the error of each predicted rating and the real rating value~\cite{liu2010unifying,qi2019time,zarzour2018new}. Nevertheless, in real-world applications, most feedback from users are presented in the implicit manner, such as clicking a product, purchasing an item, and visiting a restaurant. The implicit feedback is more common and much easier to collect than the explicit feedback. Therefore, researchers focus on the implicit feedback based recommendation.

Different from the explicit rating values, in implicit feedback recommendation scenario, we only have users' observed behaviors while the large majority of users' preferences are unknown~\cite{pan2008one,pan2013gbpr}.  As the unknown records are much larger than the observed feedbacks, some researchers proposed to use weighted regression optimization, e.g., weighted matrix factorization to assign lower weights for unobserved behaviors than the observed behaviors~\cite{lian2020personalized}. Since the task of recommendation is to provide a personalized ranking to users, a natural idea is to use ranking based optimization. The current ranking based optimization models can be mainly categorized into pairwise and listwise models. Pairwise models are based on the pairwise comparison between a relevant item and an irrelevant item at each time~\cite{rendle2012bpr,pan2013gbpr,yu2018multiple}. E.g., the most widely used pairwise method of Bayesian Personalized Ranking, assumes that a user prefers an observed item than a randomly selected unobserved item~\cite{rendle2012bpr,wu2021survey}.  Most listwise models optimize the top-N ranking oriented measure or maximize the permutation probability of the most likely permutation of the defined list, such as the overall item set~\cite{xia2008listwise,huang2015listwise}, or the list that is composed of an observed item and unobserved items~\cite{wu2018sql}. As computing top-N probability grows exponentially with N, nearly all listwise models simplify N as 1 for efficiency consideration.
These ranking based models treat the recommendation as a ranking task, and show better performance than pointwise metrics for recommendation.

In these ranking based methods, as the size of the unobserved items are much larger than the size of the observed items, researchers proposed to randomly sample a portion of items from the unobserved set as negative items at each iteration, in order to reduce time complexity without much performance loss.  Instead of the random selection, researchers proposed to select negative items based on heuristics, such as sampling items based on the popularity~\cite{wu2019noise,steck2011item,gantner2012personalized}.
Another kind of models argued that different unobserved items have different importance in the ranking process, instead of random selection or heuristic selection models, researchers designed models to dynamically choose negative training samples from the ranked list produced in the current prediction model~\cite{grbovic2018real,zhang2013optimizing,zhong2014adaptive,chen2020jointly}.  
E.g., researchers proposed a non-uniform sampling distribution that adapts to both the context and the current state of learning~\cite{rendle2014improving}.
The negative item sampling techniques further improve collaborative ranking performance by mining the valuable samples from the large set of unobserved item behaviors.

Despite the advances of collaborative ranking based approaches for implicit feedback, we argue that performance of  current ranking models for CF are still far from satisfactory. For each user, the implicit feedback are divided into two sets: an observed item set with limited observed behaviors, and a large unobserved item set that is mixed with negative and unknown behaviors. Each user's observed item size is very small, and usually far less than the size of unobserved items. E.g., the density of most user-item interaction matrix is less than 1\%. How to well exploit the structure information hidden in two sets, especially the observed set is the key challenge.
For most ranking based CF models, the limited observed set is only used at element wise level independently, while the relationships of observed items have not been well exploited.
Specifically, pairwise approaches take each observed-unobserved item pair with independence assumption of each selected pair. Listwise approaches resort to top-1 ranking for efficiency consideration and could not model observed set from a global perspective.  Given the sparse observed set and the large unobserved set, can we better learn patterns between two sets, such that more self-supervised information hidden in users' behaviors can be learned for recommendation?

In this paper, we design Set2setRank for collaborative ranking with implicit feedback. The key idea of our proposed ranking framework is that: instead of treating each element in the observed set independently, we explore how to better exploit the structure hidden in each set, and the structure information between two sets for implicit feedback.  
To this end, we design a novel Set2setRank approach to better exploit the structure information hidden in the observed set and unobserved set to guide implicit feedback based ranking. At each iteration, we  sample an observed set and an unobserved set from implicit feedback. We design a newly two-level comparison to learn more self-supervised information hidden in users' behaviors. 
The first level is an item to an item set comparison that encourages each sampled unobserved item is ranked lower than any sampled observed items. The second level is a set to set comparison to encourage a margin between distance of observed items summarized from the observed set, and distance of the most ``hard'' item selected from the first level comparison. 
Moreover, we design an adaptive sampling algorithm  to implement our proposed two-level comparison, where the size of the two sampled sets vary at each iteration. 
We have to note that our proposed framework is model-agnostic and can be easily applied to most recommendation prediction approaches, and is time efficient in practice.  Finally, extensive experiments on three real-world datasets demonstrate the superiority of our proposed approach.

\vspace{-0.1cm}
\section{Related Work}
CF exploits the collaborative signals from user-item interaction behaviors for recommendation~\cite{wu2019noise,wang2019unified,chae2020ar}. Among all models for CF, learning user and item embeddings have been popular for modern recommender systems, as embedding learning shows flexibility and relative high performance~\cite{wang2019unified,liu2019ekt}. After that, the preference of a user to an item is predicted by the inner product of the user and item embedding. Most research works have been focused on designing user and item embedding architecture. Matrix factorization based approaches learn user and item embeddding based on low rank decomposition of the user-item interaction matrix~\cite{koren2009matrix,xue2017deep}. Recently, due to the huge success of graph neural networks, researchers have treated the user-item behaviors as a user-item bipartite graph, and designed neural graph models~\cite{wang2019neural,chen2020revisiting}. E.g., NGCF is one of the first few attempts that designed embedding propagation to inject the node centric graph structure for user and item embedding learning~\cite{wang2019neural}. LightGCN~\cite{he2020lightgcn} and LR-GCCF~\cite{chen2020revisiting} use simple graph convolutions, which are easier to tune and show better performance.

Another research line of CF designed optimization and ranking goals. Earlier works modeled  explicit rating values, and most of them directly use pointwise loss of root mean squared error~\cite{qi2019time,zarzour2018new,jamali2010matrix}. As implicit feedback in more common in recommender systems, many researchers worked on implicit feedback based recommendation~\cite{yu2020sampler,lu2019effects}. In implicit feedback, there is a limited number of observed behaviors with a rating value of 1, and the remaining are unobserved behaviors. Some approaches extended pointwise approaches, and treated all unobserved items as negative items by assigning them with smaller confidence values~\cite{yu2017selection}. 

As recommendation is a ranking problem that provides top ranked items, researchers proposed  ranking based optimization approaches for recommendation. These ranking approaches can be divided into pairwise models~\cite{rendle2012bpr,pan2013gbpr} and listwise models~\cite{cao2007learning,huang2015listwise}. Bayesian Personalized Ranking~(BPR) is a popular pairwise approach that is designed for implicit feedback based recommendation~\cite{rendle2012bpr}. Given a user, by selecting an observed-unobserved item pair, BPR assumes a user prefers an observed item compared to the unobserved item. BPR is easy to implement, and the time complexity is linear with the observed ratings. However, pairwise approaches treat each item pair independently, and ignore the correlation of multiple observed items and multiple unobserved items. To explore the correlation of more data, GBPR~\cite{pan2013gbpr} and ABPR~\cite{ouyang2019asymmetric} built the correlation among users. Specifically, users with the same interests of items are grouped together, then these two models defined user group preference to build a relationship among users.
Another direction involves modeling  the correlation among multiple items~\cite{pan2013cofiset,hsieh2017collaborative}. For example, by computing the mean value of each item in a set of items to form a set preference, CoFiSet is proposed to introduce the correlation between multiple items for enhancing ranking performance~\cite{pan2013cofiset}. However, most of these methods only focus on a single aspect of correlation. How to better exploit the structure information of implicit feedback, especially with the limited observed behaviors is still under explored.

The listwise approaches formulate a listwise based ranking loss to measure the distance between predicted list and true list~\cite{xia2008listwise}, such as ListNet~\cite{cao2007learning}, ListRank-MF~\cite{shi2010list} and ListCF~\cite{huang2015listwise}. Some listwise approaches were applied for memory based CF, and needed explicit ratings for nearest neighbor ranking~\cite{huang2015listwise,wang2016ranking}. SQL-Rank is proposed to cast listwise collaborative ranking as maximum likelihood under a permutation model which applies probabilities to permutations of the item set~\cite{wu2018sql}. 
As the observed items are all treated as rating 1, SetRank is proposed to use the permutation probability to encourage  one observed item ranks in front of multiple unobserved items in each list~\cite{wang2020setrank}. SetRank provides a new research perspective for listwise learning for implicit feedback, and achieves state-of-the-art ranking results. Another direction is directly maximizing the ranking metrics~\cite{liu2015boosting,yu2020collaborative,shi2012tfmap,balakrishnan2012collaborative}. Since most of the ranking metrics are not differentiable, existing models approximate the ranking metrics and optimize a smoothed version, such as deriving the lower bound. CLiMF optimizes a lower bound of Mean Reciprocal Rank~(MRR) and becomes one of the popular listwise models for implicit feedback~\cite{shi2012climf}.  
Despite the success of existing listwise models, most of them simplified the ranking problem for top-1 recommendation with surrogate optimization goals~\cite{huang2015listwise,wang2016ranking,shi2010list,wang2020setrank}.

In implicit feedback, the data is composed of a small portion of observed behaviors and a large portion of unobserved interactions. Some algorithms sampled negative items from a large portion of unobserved interactions with equal probabilities for each user~\cite{rendle2012bpr,pan2008one,ding2018improved,wang2019variance}. Other  researchers developed techniques to pick valuable negative samples~\cite{grbovic2018real,zhang2013optimizing,zhong2014adaptive,chen2020jointly}. By combining the context like a set of users or additional variables, a non-uniform item sampler is proposed for negative sampling~\cite{rendle2014improving}. 
Instead of explicitly picking unobserved items as negative items, NCE-PLRec generates negative instances by sampling from the popularity distribution~\cite{wu2019noise}. These negative sampling methods further improve collaborative ranking performance. We also borrow the idea of negative sampling, and we explore how to better mine the relationships of positive samples from the positive set for ``hard'' negative item mining.

\begin{comment}
\begin{table}[htb] \centering
  \caption{\small{Mathematical Notations}}\label{tab:math_notations}
  \vspace{-0.2cm}
  \begin{footnotesize}
  \centering
  \begin{tabular}{|l|l|}  \hline
  Notations & Description \\ \hline \hline
  U, V &  user set, item set,~$|U|$ is number of user,\\&~$|V|$ is number of item\\ \hline 
  $\mathbf{R}=[{r}_{uv}]_{|U| \times |V|}$  & interaction matrix, with $r_{uv}$ denotes whether $u$ likes item $v$  \\ \hline
  $\hat{\mathbf{R}}=[\hat{r}_{uv}]_{|U| \times |V|}$  & predictive matrix, with $\hat{r}_{uv}$ denotes \\ & user~$u$'s predicted score to item~$v$\\ \hline
  $\mathbf{R}^+_u$  & observed feedbacks of user~$u$\\ \hline%, i.e., $r_{uv}=1$ for~$v \in \mathbf{R}^+_u$  \\ \hline
  $\mathbf{R}^-_u=\mathbf{R} \backslash\mathbf{R}^+_u$  & unobserved feedbacks of user~$u$\\ \hline%, i.e., $r_{uv}=0$ for~$v \in \mathbf{R}^-_u$  \\ \hline
  $\mathbf{S}^+_u$  & the predicted rating set of sampled observed item set,\\&~$|\mathbf{S}^+_u|=L$   \\ \hline
  $\mathbf{S}^-_u$  & the predicted rating set of sampled unobserved item set,\\&~$|\mathbf{S}^-_u|=K$  \\ \hline 
  \end{tabular}
  \end{footnotesize}
  \vspace{-0.2cm}
\end{table}
\end{comment}

\begin{small}
  \begin{figure*} [htb]
    \begin{center}
      \includegraphics[width=140mm]{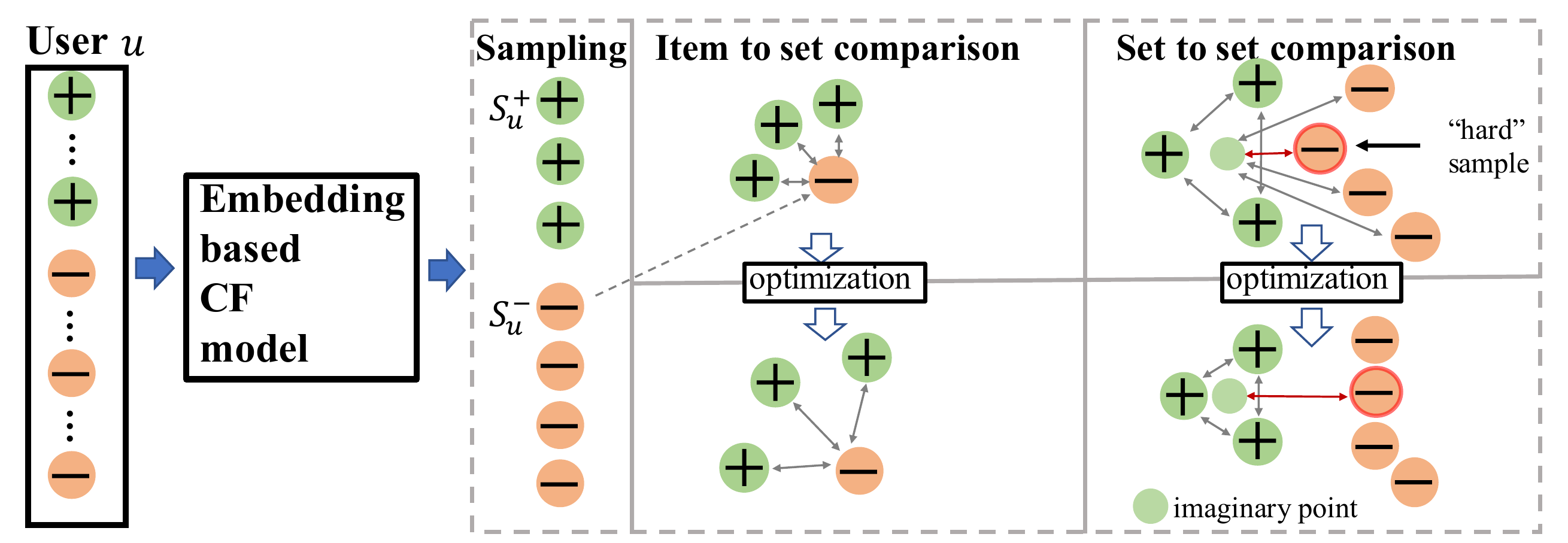}
    \end{center}
       \vspace{-0.5cm}
     \caption{The overall structure of our proposed framework. After sampling observed set~$S_u^+$ and unobserved set~$S_u^-$, we build two-level comparisons that encourage the ranking positions of observed set before unobserved set. The item to set comparison pushes each unobserved sample far from observed set. The set to set comparison assumes that the summarized distance among observed set should be smaller than the distance between observed samples and the most ``hard'' unobserved sample.}\label{fig:overall structure}
      \vspace{-0.3cm}
  \end{figure*}
\end{small}

\section{Preliminary}
Now, we define a general recommendation problem.
Let  user set~$\mathbf{U}$ and item set~$\mathbf{V}$ denote two kinds of entities of a recommender system, with $|U|$ and~$|V|$ are the size of~$\mathbf{U}$ and~$\mathbf{V}$.
Let~$\mathbf{R}=[r_{uv}]_{|U|\times |V|} \in\{0,1\}$ denote the user-item interaction matrix,  with each element~$r_{uv}$ in~$\mathbf{R}$ represents the implicit preference of user~$u$ to item~$v$.~$r_{uv}=1$ indicates the observed feedback~(i.e., user~$u$ likes the item~$v$),~$r_{uv}=0$ indicates the unobserved feedback~(i.e., the interaction between $u$ and $v$ is unobserved).
For each user~$u \in \mathbf{U}$, we use~$R^+_u$ to indicate observed feedback, by contrast,~$R^-_u=\mathbf{R} \backslash R^+_u$ to indicate non-interactive feedback. We have $R^+_u\cap R^-_u=\emptyset$ and  $R^+_u\cup R^-_u=\mathbf{V}$. Given the user interaction matrix~$\mathbf{R}$, the goal of CF ranking is to identify user's preferences and predict the preference score matrix~$\hat{\mathbf{R}}=[\hat{r}_{uv}]_{|U|\times |V|}$ with ranking based optimization functions.

We focus on design a ranking approach to predict~$\hat{\mathbf{R}}$. As embedding based models have been the default choice of recommendation architecture, we can use any embedding based CF model to learn users'  and items' embeddings~$\mathbf{E}=\{\mathbf{E}_{U},\mathbf{E}_{V}\}$. Specifically, $\mathbf{E}_{U}=\{e_1,...,e_u,...,e_{|U|}\}$ is the user embeddings, and ~$\mathbf{E}_{V}=\{e_{|U|+1},...,\\e_v,...,e_{|U|+|V|}\}$ is the item embeddings. After that, we can predict each user $u$'s preference to item $v$ as~$\hat{r}_{uv}=e_u^Te_v$.
% \begin{small}
% \begin{equation}
% \hat{r}_{uv}=e_u^Te_v.
% \end{equation}
% \end{small}
Please note that as we focus on ranking based optimization design, we assume the architecture of embedding learning is available, such as matrix factorization models~\cite{rendle2012bpr} or neural graph models~\cite{chen2020revisiting}.

\subsection{Pairwise and Listwise Learning}
In order to predict users' preferences, we need to use information of implicit feedback. In common implicit feedback, the mainly self-supervised information is a kind of preference assumption between observed and unobserved feedback, such as observed feedback $>_u$ unobserved feedback where $>_u$ represents the preference structure of user~$u$. Pairwise learning is one of the most widely techniques to define a pairwise preference assumption. For each user $u$, pairwise learning assumes and uses self-information that the connected observed item~$i$ is more relevant than sampling unobserved item~$j$, which can be formulated as follows:
\begin{flalign}
  i >_u j, \quad \forall i \in R_u^+, \quad \forall j \in R_u^-.
%   \text{observed item} >_u \text{unobserved item},
\end{flalign}
To this end, the corresponding ratings can be formulated as follows:
\begin{flalign}
\hat{r}_{ui}>\hat{r}_{uj},
\end{flalign}
where~$i \in R^+_u$ and~$j \in R^-_u$.
In this situation, each pairwise item preference data is modeled independently without considering the complex correlations of multiple items~\cite{wang2020setrank}. 
For example, a user $u$ likes item $i_1$ and $i_2$, and has an unobserved item set of $[j_1,j_2,j_3]$. By independently sampling two pairwise relations: ~$i_1 >_u j_1$ and~$i_2 >_u j_2$, the ground truth of $i_2 >_u j_1$ is not considered due to the independence assumption. Thus, the self-supervised information of implicit feedback is not fully utilized.

Listwise learning directly looks at all items~\cite{wu2018sql} or entire observed items~\cite{shi2012climf} at one time, and assumes that the correct sorting is better than the incorrect sorting.  
Let~$\pi$ be a ranking permutation~(or sorting) of a list of items for a user. With the predicted ratings of each ranked user-item pair as $\hat{r}_{\pi_{j}}$, the permutation probability can be calculated as:
$P(\pi)=\prod_{j=1}^{|\pi|} \frac{\phi\left(\hat{r}_{\pi_{j}}\right)}{\sum_{l=j}^{|\pi|} \phi\left(\hat{r}_{\pi_{l}}\right)}$, where$\phi()$ is a monotonically increasing function that transforms the predicted rating into probability. Then, the goal of the listwise ranking is to find the best permutation probability of all users given the ground-truth rankings from the implicit feedback.  
In fact, this permutation probability needs to be calculated every time when the list has changed its order. Due to the high time cost of permutation probability calculation, nearly all listwise models consider top-1 for simplicity. However, this simplification ignores most of the information hidden in the entire observed item set, which leads to insufficient usage of implicit feedback information.

It is important to understand that these two ways~(pairwise learning and listwise learning) do not make full use of implicit feedback. Given the uniqueness of the implicit feedback based CF, the following question should be considered: since observed items and unobserved items are disordered inside each set, sampling entire items and sorting them is time consuming and unnecessary. On the other hand, considering only one positive item is insufficient for the utilization of limited observed items. Therefore, how to explore the self-supervised relationship for implicit feedback based ranking is the main challenge that we need to tackle.

\section{The Proposed Set2setRank Framework}
In this section, we propose Set2setRank framework.   In order to make full use of the implicit feedback, we propose to analyze the interactions between items and users at set level. Specifically, we first sample an observed set and  an unobserved set. Then, we introduce the optimization and two-level comparison of our proposed Set2setRank framework. 
For easier understanding, Figure.~\ref{fig:overall structure} illustrates the overall architecture of our proposed Set2setRank framework with the two-level comparison. After that, we design an adaptive sampling technique to further improve the ability of Set2setRank framework. Finally, we make a discussion about the properties of our proposed Set2setRank framework.

\subsection{Construction of Two Sets} 
As mentioned before, pairwise method suffers from independent assumption. 
Meanwhile, comparing all items in listwise methods leads to low efficiency problem. To this end, we try to make a compromise between these two methods. Specifically, we focus on an intermediate sampling to flexible use of observed and unobserved feedback. 
We intend to sample $L$ observed items from the interactive item set $R_{u}^+$, and utilize the corresponding predictive score to construct the observed set $S_{u}^+$ during training. Meanwhile, the same operation is applied to construct the unobserved set $S_{u}^-$.  
This process can be formulated as follows:
\vspace{-0.1cm}
\begin{equation}
    \begin{split}
        % S_{u}^+ &= \{\hat{r}_{ui}^+ \in \hat{R}|i=1,2,...,L, r_u^i\inR_{u}^+\}, \\
        % S_{u-} &= \{\hat{r}_{u-}^j \in \hat{R}|j=1,2,...,K, r_u^j\inR_{u-}\},\\
        S_u^+ &= \{\hat{r}_{ui}^+|i\in R_{u}^+\}, \quad |S_u^+| = L, \\
        S_u^- &= \{\hat{r}_{uj}^-|j\in R_{u}^-\}, \quad |S_u^-| = K, \\
        % S_u^+ &= \{i|r_{ui}\in R_{u}^+\}, \quad |S_u^+| = L, \\
        % S_u^- &= \{j|r_{uj}\in\mathbf{R}_{u}^-\}, \quad |S_u^-| = K, \\
    \end{split}
\end{equation}
% \vspace{-0.1cm}
where $i$ and $j$ indicate the $i^{th}$ observed sample and the $j^{th}$ unobserved sample for user $u$.% sampled from the observe set $R_{u}^+$ and unobserved set $R_{u}^-$, respectively.
For user~$u$, $\hat{r}_{ui}^+$ and $\hat{r}_{uj}^-$ denote the predicted ratings for the sampled observed item $i$ and sampled unobserved item $j$. 
$K$ and $L$ are the sample size of $S_u^+$ and $S_u^-$. 
Since the size of observed items is much smaller than unobserved items, comparing them at set level can not only fully exploit the structure information hidden in each set, but also effectively analyze the noise unobserved information with the guidance of observed information.

\subsection{Optimization Criteria of Set2setRank}
Since we have sampled two sets of each user, we aim at maximizing  the following preference:
\vspace{-0.1cm}
\begin{flalign}\label{eq:set_cmp_all} 
  S_u^+ >_u S_u^-,
\end{flalign}
\vspace{-0.1cm}
where $>_u$ denotes the preference structure of user~$u$. In order to model the interactions of users and items based on the observed set and unobserved set in a comprehensive way, we design two-level comparisons~(i.e. \textit{item to set comparison} and \textit{set to set comparison}) to achieve this constraint~(Eq.\eqref{eq:set_cmp_all}) concretely.
In the following part, we first introduce the preference gain. Then, we give a detailed analysis about our newly designed two-level comparison.

% \subsubsection{Distance Function and Comparison Function}
\subsubsection{Preference Gain Function and Comparison Function}

Eq.\eqref{eq:set_cmp_all} describes the preference of user $u$, in which the observed items should be rated higher than the unobserved items. One step further, we should rank the observed items in front of the unobserved items. However, Eq.\eqref{eq:set_cmp_all} is not differentiable. 
We need a replacement to satisfy the requirements above and also be differentiable. 
To this end, we define a preference gain~$D(x,y)$, where $x$ and $y$ are sampled from predictions of the observed set~$S_u^+$ and predictions of unobserved set~$S_u^-$, separately. The output should be large, indicating the predicted ratings of the two samples  are far away. 
Since the pairwise loss function in BPR is an intuitive and widely used preference function in implicit feedback, we adopt the same preference metric as BPR~\cite{rendle2012bpr}, which is formulated as follows:
\vspace{-0.1cm}
\begin{flalign}\label{eq:distance_cal}  
  D(x, y) = \sigma(x-y),
\end{flalign}
% \vspace{-0.1cm}
where $\sigma()$ is the activation function. $\{x, y\}$ is the input of this function, and is two scalar data like two samples from~$S_{u}^+$ or~$S_{u}^-$. 
 
For set comparison, we intend to fully exploit the structure information hidden in the implicit feedback and make full use of observed set to model user preference. 
Therefore, we try to build multiple comparisons/views between observed set and unobserved set. More specifically, we need each sample in one set (e.g., $S_{u}^+$) to be far away from the entire another set (e.g., $S_{u}^-$). 
To this end, we define a comparison function for our proposed set comparison on the set level as follows:
% \vspace{-0.1cm}
\begin{equation} 
\label{eq:function_f} 
    F(X,y_{j}) = \sum_{i=1}^{|X|} [D(x_i,y_j)],
\end{equation}
% \vspace{-0.1cm}
where $y_j$ denotes the $j^{th}$ elements in the set~$Y$ (e.g., unobserved set $S_{u}^-$), and $X$ represents another set (e.g., observed set $S_{u}^+$).~$|X|$ is the number of elements in the set~$X$. Next, we introduce the technical details of our proposed two-level comparison on the sets.

\subsubsection{Item to Set Comparison}
Different from the comparison of a pair of items at each time of pairwise learning models, the comparison between observed set and unobserved set has to deal with multiple samples. Therefore, how to effectively and fully use set data to ensure the personalized recommendation performance is our main focus. There is explicit set-level preference information. For one user, we have  confidence that most unobserved samples have lower predicted ratings than ratings of the observed set.  
In other words, the predicted rating of each unobserved item is encouraged to be smaller than all the predicted ratings of observed items. This constraint can be implemented as follows:
\vspace{-0.1cm}
\begin{equation}
\label{eq:npair_loss}
    \begin{split}
        L_2(u) = \underset{{S}_{u}^+}{\mathbb{E}}\underset{{S}_{u}^-}{\mathbb{E}}[\sum_{j=1}^K ln~F(S_{u}^+,\hat{r}_{uj}^-)],
        % [\sum_{i=1}^LF(\hat{r}_{ui}^+, S_{u}^-) + \sum_{j=1}^K ln F(\hat{r}_{uj}^-, S_{u}^+)].
    \end{split}
\end{equation}
where~$\{\hat{r}_{uj}^-|j=1,...,K\}$ is all the samples in~$S_u^-$. %Similar as many works for distance measures~\cite{rendle2012bpr}, we use~$ln$ function to expand the distance.

By utilizing this loss function, the recommendation method can push the distance between observed samples(set) and unobserved set(samples) larger. Please note that, this formulation differs from pairwise models, as pairwise models can only push two items far away at one time. In contrast, the item to set level comparison pushes each unobserved item far away from entire items in the observed set for optimization.

\subsubsection{Set to Set Comparison}
\label{s:set2set_compare}
The item to set comparison has exploited the structure information hidden in implicit feedback by optimizing 
$L+1$ items~(one negative item and $L$ observed items from the sampled observed set) at one optimization step. Since the size of observed items is much smaller than unobserved items in the implicit feedback, item to set comparison still is in short of fully utilizing the structure information in observed set.
In view of this drawback, we propose a set to set comparison to fully explore observed set and exploit useful information from the noise unobserved set. 
We assume that the summarized distance among observed set should be smaller than the distance between observed samples and the most ``hard'' unobserved sample.%~(i.e., the unobserved sample that is nearest to the observed set).

Specifically, for each element $\hat{r}_{ui}^+$ in the observed set, the summary of an observed set~$S_u^+$ can be implemented by directly measuring the distance between observed samples:
\vspace{-0.1cm}
\begin{flalign}\label{eq:pos_summary}  
% f_{pos}(S_{u}^+, S_{u}^+) = ln\sum_{i=1}^{L} ~F(S_{u}^+,\hat{r}_{ui}^+).
f_{pos}(S_{u}^+, S_{u}^+) = 1/L\sum_{i=1}^{L} ln~F(S_{u}^+,\hat{r}_{ui}^+).
\end{flalign}
\vspace{-0.15cm}
 
For each item from the unobserved set $S_u^-$, since we have already calculated the distance between the observed set and each unobserved item in Eq.\eqref{eq:npair_loss}, we pick the ``hard'' unobserved item as: 
\vspace{-0.1cm}
\begin{equation}\label{eq:neg_summary}
\begin{split}
    f_{neg}({S}_{u}^+,{S}_{u}^-) &= min([ln~F( S_{u}^+,\hat{r}_{u1}^-), ln~F( S_{u}^+,\hat{r}_{u2}^-),...,ln~F(S_{u}^+,\hat{r}_{uK}^-)]) \\
    &= min([ln~F( S_{u}^+,\hat{r}_{uj}^-)|j=1,...,K]),
\end{split}
\end{equation}
\vspace{-0.1cm}
where~$S_u^-=\{\hat{r}_{u1}^-,...,\hat{r}_{uK}^-\}$,~$min$ represents picking the most ``hard" unobserved item that is closely to the observed set. It combines the idea of negative sampling and has the advantage of low computational complexity.

As the distance of observed items~(Eq.\eqref{eq:pos_summary}) is smaller than the distance of the most ``hard'' unobserved items, we enlarge the margin:
\vspace{-0.3cm}
\begin{flalign}\label{eq:set_loss}
L_3(u) = \underset{{S}_{u}^+}{\mathbb{E}}\underset{{S}_{u}^-}{\mathbb{E}} ln~D(f_{neg}({S}_{u}^+,{S}_{u}^-),\beta * f_{pos}({S}_{u}^+, {S}_{u}^+)),
\end{flalign}
% \vspace{-0.1cm}
% \\ \nonumber&
where~$\beta$ is a margin parameter to adjust the distance between sets.

Another intuitive idea is that, similar as selecting the ``hard''  unobserved sample to keep the set-level distance, we can also select the ``easy'' observed sample for set to set comparison. Then,  we push the ``easy'' observed sample far from the ``hard'' unobserved sample. This process can be achieved as:
% through the following calculations:
\vspace{-0.1cm}
\begin{equation}\label{eq:easy_summary} 
    \begin{split}
        g_{pos}({S}_{u}^+) &= max([ln~F(S_{u}^+,\hat{r}_{ui}^+)|j=1,2,...,L]), \\%,S_u^+=\{\hat{r}_{u1}^+,...,\hat{r}_{uL}^+\}
        L_4(u) = & \underset{{S}_{u}^+}{\mathbb{E}}\underset{{S}_{u}^-}{\mathbb{E}} ln~D(f_{neg}({S}_{u}^+,{S}_{u}^-),\beta *g_{pos}({S}_{u}^+)).
    \end{split}
\end{equation}
In practice, we find that ~$g_{pos}(S_u^+)$ performs a little worse than 
the performance as~${f}_{pos}({S}_{u}^+,S_u^+)$. A possible reason is that, 
the ``easy '' observed positive item has already been calculated in the item to set level comparison. By selecting the ``easy '' observed positive item, we only increase the weight of a single observed item.
In the following, we select~$f_{pos}(S_u^+,S_u^+)$ to achieve the set-level constraint.
%  relative distance

\subsubsection{Objective Function}
We integrate two set comparison losses to obtain two correlated views of the same user. The overall loss can be formulated as maximizing the following loss:
\begin{flalign} \label{eq:overall_loss}
Loss = \sum_{u\in U} Loss(u) = \sum_{u\in U}[L_2(u)+\lambda*L_3(u)],
\end{flalign} 
where~$\lambda$ is a weighting parameter. By taking the consideration of two-level comparison, $Loss(u)$ can force Set2setRank to make observed items closer to the user and unobserved items far away from the user, which is in favor of providing better results.

\subsection{Adaptive Set2setRank} ~\label{subsec: aset2setrank}
In the previous part, we require each observed set have $L$ items. In fact, the number of observed items varies greatly among different users, with some users have very large number of interaction records, while others have only  several ratings~\cite{chen2020esam,shi2020beyond}.  Therefore, instead of a fixed size $L$ of the sampled observed set, the size of the observed set can be flexibly designed. To further enhance the ability of our proposed framework, we design a simple and effective mask method, which can
flexibly select the set size.

Word masking randomly picks and replaces some words, which is widely applied in representation learning in many natural language processing scenarios~\cite{devlin2018bert,cui2019pre}. Inspired by word masking, we enhance Set2setRank in a similar way. Specifically, in each observed set~$S_u^+$, we use a random mask to remove some observed samples and keep at least two observed samples. 
The mask~${Mask}=\{{Mask}_1,...,{Mask}_l,...,$ ${Mask}_L\}$ randomly selects some observed samples from observed set in each update. In order to integrate the ${Mask}$ into our framework, we modify Eq.\eqref{eq:function_f} with the~${Mask}$ as follows:
% \begin{small}
\begin{flalign} 
{Mask}_l &=\left\{\begin{array}{ll}
1, & selected \\
0, & \text { else }
\end{array}\right.
\text{and} \sum_{l=1}^L {Mask}_{l} \geq 2, \\
\tilde{F}(X,y_{j}) &= \sum_{i=1}^{|X|} [D(x_i,y_j)*Mask_i],
\end{flalign}
% \end{small}
please note that,~$Mask$ is generated by a random probability. 

We have to note that our proposed $Mask$ is compositional, which can be flexibly formed with different combinations of set lengths during training. This is equivalent to using a variety of sampling methods to form the observed set, thus observed set is actually changeable and dynamic to suit different lengths. Therefore, using the mask method can adaptively deal with the different numbers of observed samples. In this way, the item to set comparison~$L_2$ in Eq.\eqref{eq:npair_loss} will be modified as follows:
\vspace{-0.1cm}
\begin{flalign} \label{eq:newl2_loss} 
  L_2(u) = \underset{{S}_{u}^+}{\mathbb{E}}\underset{{S}_{u}^-}{\mathbb{E}}[\sum_{j=1}^K ln~\tilde{F}(S_{u}^+,\hat{r}_{uj}^-)].
\end{flalign}
Meanwhile, Eq.\eqref{eq:pos_summary} and Eq.\eqref{eq:neg_summary} will be changed into:
\begin{flalign} \label{eq:new_summary}
  \tilde{f}_{pos}(S_{u}^+,S_{u}^+) &= 1/L \sum_{i=1}^{L}ln\tilde{F}(S_{u}^+,\hat{r}_{ui}^+),\\ 
  \tilde{f}_{neg}(S_{u}^+, S_{u}^-)&= min([ln~\tilde{F}(S_{u}^+,\hat{r}_{uj}^-)|j=1,2,...,K]).
\end{flalign}

By applying mask technology into Set2setRank, our proposed framework can deal with  different numbers of observed samples for each user. Along this line, it can flexibly accommodate different combinations of observed samples. Thus, it is able to cope with different interactions and improves its generalization ability.

\subsection{Model Discussion}

\textbf{Connections with Previous Works.} 
In order to make full use of implicit feedback, we propose to compare the observed interactions and unobserved interactions at set level. This is the main difference of Set2setRank compared with pairwise and listwise methods.
By introducing the item to set level comparison, we break the independence assumption of each item pair in pairwise based methods, and can jointly compare multiple unobserved items at the same time. Besides, the set to set level comparison pushes away the distance of positive items, and distance of the ``hard'' negative items picked from the item to set level calculation. The set to set level comparison jointly considers observed items, unobserved items together by modeling self-supervised information of the limited observed behaviors. However, listwise models either considered top-1 recommendation or surrogate optimization function with efficiency consideration, and could not well exploit the limited observed behavior for better ranking.

In practice, if we simply set~$\lambda=0$ and~$|S^+_u|=|S^-_u|=1$ in Eq.\eqref{eq:overall_loss}, $Loss(u)$ degenerates to pairwise loss, which is a typical target of pairwise learning. 
Furthermore, if we enlarge $|{S}^+_u|$ and~$|{S}^-_u|$ to maximum lengths, Set2setRank is modified to a special listwise method, where a list of all observed items and all unobserved items are considered for ranking. 
%  (i.e., the size of all the data of training data)

\textbf{Complexity Analysis.}
We perform complexity analysis in this part. First of all, we introduce some notations.  Let ~${M}_{avg}$ denote the average number of user's observed items,~$M_{max}$ denote the maximum number of users' observed items. As each user has very limited observed behaviors, e.g., the density of most CF data is less than 1\%, the average number of unobserved items can be regarded as the item set size.
Therefore, almost all methods use sampling  techniques to pick some unobserved items, and sample~$K$ unobserved items at a time. 
We analyze our proposed framework with four typical ranking methods, including pairwise ranking model of BPR~\cite{rendle2012bpr}, listwise ranking models of CLiMF~\cite{shi2012climf} and SetRank~\cite{pang2020setrank}, and a negative sampling based approach of AoBPR~\cite{rendle2014improving}. As all of these models are designed for ranking based optimization, for fair comparison, we omit the time complexity of predicting users' preferences, and only compare the time complexity of the loss optimization part.
\begin{itemize}
  \item BPR~\cite{rendle2012bpr}: BPR compares one observed item and one unobserved item at each time,  and the total number of unobserved items is~$|U|M_{avg}K$. Thus, its time complexity is ~$O(|U|M_{avg}K)$.
  \item CLiMF~\cite{shi2012climf}: It compares the relevance scores between all the observed items to optimize a smoothed version of the MRR. As the number of observed items vary largely of different users, the time complexity depends on the users that have largest observed set size.
  Therefore, its time complexity can be calculated as $O(|U|M_{max}^2)$.
  \item SetRank~\cite{wang2020setrank}: SetRank compares each observed item and~$K$ unobserved items to form a list of $K+1$ items with permutation probability calculation. For each user, SetRank needs to formulates $M_{max}$ sets to ensure each observed item is considered. Its time complexity is $O(|U|M_{max}(K+1))$.
  \item AoBPR~\cite{rendle2014improving}: It is an adaptive sampling model based on BPR. The time complexity of adaptive sampling is~$O(D+|V|log|V|)$, where~$D$ is the dimension of the embedding vector. Therefore, the overall time complexity is the time complexity of BPR plus the adaptive sampling, which equals to~$O(|U|M_{avg}K(D+|V|log|V|))$.
 
\end{itemize}
For Set2setRank, the time cost mainly depends on the complexity of item to set comparison~(Eq.\eqref{eq:npair_loss}) and set to set comparison~(Eq.~\eqref{eq:set_loss}). For each user, Set2setRank samples~$L$ observed items and~$K$ unobserved items at each time.
Since each user's observed items need to be considered, all observed rating records ($|U|M_{avg}$) need to be sampled~$|U|M_{avg}/L$ times. For item to set comparison~(Eq.\eqref{eq:npair_loss}), it involves~$KL$ comparisons of~$L$ observed items and~$K$ unobserved items. Thus, its time complexity can be calculated as~$O((|U|M_{avg}/L)KL)=O(|U|M_{avg}K)$. The~$KL$ comparison in Eq.\eqref{eq:npair_loss} can be reused in Eq.\eqref{eq:neg_summary}. The additional time complexity in set to set comparison~(Eq.\eqref{eq:set_loss}) only comes from~$f_{pos}$~(Eq.\eqref{eq:pos_summary}). The~$f_{pos}$ involves~$L^2$ comparisons, its time complexity can be calculated as~$O((|U|{M}_{avg}/L)L^2)=O(|U|{M}_{avg}L)$. Therefore, the overall time complexity is~$O(|U|{M}_{avg}(K+L))$.  

For most recommender systems, ${M}_{max}>>{M}_{avg}$ always exists  (e.g.,~${M}_{max}=6479$ and~${M}_{avg}=101$ on MovieLens-20M dataset). Thus, two listwise approaches~(CLiMF and SetRank) have the higher time complexity. For BPR and Set2setRank, their time complexity are very similar. In practice, for our proposed Set2setRank, the values of $K$ and $L$ are very small, and we find $L=K-1$ achieves relatively high ranking performance. Thus, we have~$O(|U|{M}_{avg}(K+L))=O(|U|{M}_{avg}K)$. The above time complexity of Set2setRank is very similar to the time complexity of BPR.

% \begin{table}[]
% \caption{\small{The statistics of the datasets.}}\label{tab:data_stats} 
% \begin{scriptsize}
% \begin{tabular}{|l|l|l|l|l|}
% \hline
% Dataset & Users & Items & Ratings & Rating Density \\ \hline
% MovieLens-20M & 138,493 & 26,744 & 20,000,263 & 0.54\% \\ \hline
% Amazon-book & 52,643 & 91,599 & 2,984,108 & 0.062\% \\ \hline
% Yelp & 45,919 & 45,538 & 1, 185, 068 & 0.057\% \\ \hline
% \end{tabular}
% \end{scriptsize}
% \end{table}

\begin{table*}[]
\vspace{-0.5cm}
\footnotesize
\centering
\caption{Overall performance of HR@N and NDCG@N on three datasets. }\label{tab:resall} 
\vspace{-0.4cm}
  \begin{tabular}{|l|l|l|l|l|l|l|l|l|l|l|l|}
  \hline
  \multirow{2}{*}{Dataset} & \multicolumn{1}{c|}{\multirow{2}{*}{Models}} & \multicolumn{2}{c|}{N=10} & \multicolumn{2}{c|}{N=20} & \multicolumn{2}{c|}{N=30} & \multicolumn{2}{c|}{N=40} & \multicolumn{2}{c|}{N=50} \\ \cline{3-12} 
   & \multicolumn{1}{c|}{} & HR & NDCG & HR & NDCG & HR & NDCG & HR & NDCG & HR & NDCG \\ \hline
  \multirow{10}{*}{AmazonBooks} & BPR & 0.01851 & 0.01710 & 0.02853 & 0.02169 & 0.03821 & 0.02564 & 0.04737 & 0.02911 & 0.05556 & 0.03205 \\ \cline{2-12} 
   & Pop-Sampling & 0.01945 & 0.01774 & 0.03028 & 0.02268 & 0.04072 & 0.02693 & 0.05033 & 0.03056 & 0.05914 & 0.03373 \\ \cline{2-12} 
   & AoBPR & 0.01960 & 0.01787 & 0.03061 & 0.02288 & 0.04157 & 0.02735 & 0.05123 & 0.03101 & 0.06019 & 0.03423 \\ \cline{2-12} 
   & CLiMF & 0.01546 & 0.01431 & 0.02351 & 0.01790 & 0.03079 & 0.02086 & 0.03761 & 0.02343 & 0.04392 & 0.02568 \\ \cline{2-12} 
   & Deep-SetRank & 0.01950 & 0.01827 & 0.02913 & 0.02259 & 0.03879 & 0.02650 & 0.04751 & 0.02980 & 0.05592 & 0.03282 \\ \cline{2-12} 
   & Set2setRank\_BPR & 0.02084 & 0.01900 & 0.03221 & 0.02417 & 0.04276 & 0.02846 & 0.05288 & 0.03228 & 0.06189 & 0.03550 \\ \cline{2-12} 
   & Set2setRank(A)\_BPR & \textbf{0.02169} & \textbf{0.01985} & \textbf{0.03343} & \textbf{0.02520} & \textbf{0.04446} & \textbf{0.02968} & \textbf{0.05441} &\textbf{ 0.03345} & \textbf{0.06359 }& \textbf{0.03675} \\ \cline{2-12} 
   & LR-GCCF &{0.02209} & {0.02040} & {0.03407} & {0.02583} & {0.04532} & {0.03039} & {0.05532} & {0.03416} & {0.06498} & {0.03761} \\ \cline{2-12} 
   & Set2setRank\_GCN & 0.02247 & 0.02079 & 0.03454 & 0.02629 & 0.04567 & 0.03089 & 0.05648 & 0.03490 & 0.06597 & 0.03830 \\ \cline{2-12} 
   & Set2setRank(A)\_GCN & \textbf{0.02264 }& \textbf{0.02086} & \textbf{0.03453 }& \textbf{0.02626} & \textbf{0.04598 }& \textbf{0.03092} & \textbf{0.05658 }& \textbf{0.03492} & \textbf{0.06620} & \textbf{0.03838} \\ \hline
   \hline
   \multirow{10}{*}{MovieLens-20M} & BPR & 0.2380 & 0.2504 & 0.2512 & 0.2464 & 0.2767 & 0.2522 & 0.3012 & 0.2593 & 0.3240 & 0.2668 \\ \cline{2-12}
   & Pop-Sampling & 0.2545 & 0.2707 & 0.2667 & 0.2654 & 0.2935 & 0.2714 & 0.3195 & 0.2791 & 0.3428 & 0.2871 \\ \cline{2-12}
   & AoBPR & 0.2381 & 0.2523 & 0.2522 & 0.2487 & 0.2784 & 0.2547 & 0.3027 & 0.2618 & 0.3254 & 0.2683 \\ \cline{2-12} 
   & CLiMF & 0.1130 & 0.1268 & 0.1113 & 0.1174 & 0.1185 & 0.1163 & 0.1279 & 0.1178 & 0.1373 & 0.1202 \\ \cline{2-12}
   & Deep-SetRank & 0.2759 & 0.2857 & 0.3010 & 0.2880 & 0.3314 & 0.2966 & 0.3587 & 0.3056 & 0.3841 & 0.3144 \\ \cline{2-12}%s7 new server 7
   & Set2setRank\_BPR & 0.2899 & 0.3047 & 0.3046 & 0.3005 & 0.3333 & 0.3077 & 0.3623 & 0.3168 & 0.3886 & 0.3257  \\ \cline{2-12}
   % Set2setRank\_BPR & 0.3072 & 0.3234 & 0.3245 & 0.3195 & 0.3539 & 0.3266 & 0.3838 & 0.3360 & 0.4112 & 0.3453 \\ \hline 
   & Set2setRank(A)\_BPR & \textbf{0.3067} & \textbf{0.3197} & \textbf{0.3258} & \textbf{0.3176} & \textbf{0.3567} & \textbf{0.3255} & \textbf{0.3869} & \textbf{0.3355} & \textbf{0.4144} & \textbf{0.3445} \\ \cline{2-12} 
   & LR-GCCF & 0.2672 & 0.2764 & 0.2832 & 0.2752 & 0.3125 & 0.2837 & 0.3426 & 0.2929 & 0.3715 & 0.3042 \\ \cline{2-12} 
   & Set2setRank\_GCN  & 0.2689 & 0.2801 & 0.2832 & 0.2771 & 0.3114 & 0.2847 & 0.3403 & 0.2942 & 0.3675 & 0.3036 \\ \cline{2-12}
   % Set2setRank\_GCN & 0.2735 & 0.2877 & 0.2876 & 0.2837 & 0.3162 & 0.2910 & 0.3447 & 0.3001 & 0.3714 & 0.3092 \\ \hline  
   & Set2setRank(A)\_GCN & \textbf{0.2920} & \textbf{0.3011} & \textbf{0.3135} & \textbf{0.3015} & \textbf{0.3470} & \textbf{0.3111} & \textbf{0.3805} & \textbf{0.3222} & \textbf{0.4115} & \textbf{0.3329} \\ \hline%lrgccfnewm7e90
   \hline
   \multirow{10}{*}{Yelp} & BPR & 0.03708 & 0.02929 & 0.05952 & 0.03786 & 0.07788 & 0.04406 & 0.09523 & 0.04942 & 0.11064 & 0.0539 \\ \cline{2-12}%s6,105
   & Pop-Sampling & 0.03580 & 0.02806 & 0.05655 & 0.03603 & 0.07527 & 0.04229 & 0.09116 & 0.04721 & 0.1047 & 0.05118 \\ \cline{2-12}%s13e102 
   & AoBPR & 0.03788 & 0.02951 & 0.06024 & 0.03808 & 0.08006 & 0.04469 & 0.09762 & 0.05015 & 0.11348 & 0.05478 \\ \cline{2-12}%adapt1e177
   & CLiMF & 0.01752 & 0.01329 & 0.03043 & 0.01828 & 0.04241 & 0.02235 & 0.05341 & 0.02576 & 0.06408 & 0.02887 \\ \cline{2-12}%climfs3e75
   & Deep-SetRank & 0.03801 & 0.02945 & 0.06242 & 0.03880 & 0.08353 & 0.04589 & 0.10219 & 0.05164 & 0.11886 & 0.05649 \\ \cline{2-12} %s0e117，newserver
   % BPR-fix(L2) & 0.03833 & 0.03018 & 0.06143 & 0.03905 & 0.08173 & 0.04586 & 0.09876 & 0.05113 & 0.11545 & 0.05597 \\ \hline %onlypairs19e330 
   % BPR-fix(L3) &0.03810 & 0.3012 & 0.06045 & 0.03869 & 0.08001 & 0.04526 & 0.09852 & 0.05098 & 0.11430 & 0.05558 \\ \hline%onlysets4e110
   & Set2setRank\_BPR& 0.03855 & 0.03054 & 0.06120 & 0.03918 & 0.08161 & 0.04604 & 0.09939 & 0.05153 & 0.11532 & 0.05618 \\ \cline{2-12}%fixs7e140
   % BPR-adapt(L=4,K=5)
   & Set2setRank(A)\_BPR & \textbf{0.03920} & \textbf{0.03104 }& \textbf{0.06289 }& \textbf{0.04010 }& \textbf{0.08393} & \textbf{0.04716} & \textbf{0.10228} & \textbf{0.05283} & \textbf{0.11924} & \textbf{0.05778} \\ \cline{2-12}%adapts2e280
   & LR-GCCF &0.03919 & 0.03118 & 0.06210 & 0.03991 & 0.08224 & 0.04663 & 0.10078 & 0.05236 & 0.11755 & 0.05725 \\ \cline{2-12}%lrgccfs0e100
   % LR-GCCF-fix(L=2,K=5)
   & Set2setRank\_GCN& 0.03990 & 0.03180 & 0.06238 & 0.04036 & 0.08300 & 0.04726 & 0.10092 & 0.05279 & 0.11710 & 0.05748 \\ \cline{2-12}%fixlrgccfs1e150
   & Set2setRank(A)\_GCN & \textbf{0.04058} & \textbf{0.03258} & \textbf{0.06421} & \textbf{0.04162} & \textbf{0.08509} & \textbf{0.04864} & \textbf{0.10418} & \textbf{0.05454} & \textbf{0.12070} & \textbf{0.05937} \\ \hline%adaptlrgccfs1e180

  \end{tabular}
\vspace{-0.2cm}
\end{table*}

\section{Experiments}
\label{s:experiment}

\subsection{Experimental Setup}

\textbf{Datasets and Evaluation Metrics.} 
We use three publicly available datasets: {Amazon Books\footnote{\url{http://jmcauley.ucsd.edu/data/amazon}.}}, {MovieLens-20M\footnote{\url{https://grouplens.org/datasets/movielens/20m}.}} and {Yelp\footnote{\url{https://www.yelp.com/dataset}.}}. %Table~\ref{tab:data_stats} summarizes the detailed statistics of three datasets. 
We follow the setting of Amazon Books in previous works~\cite{chen2020revisiting,wang2019neural}, Amazon Books provides 3 million ratings from 52 thousand users on 91 thousand items. Yelp dataset is adopted from the 2018 edition of the Yelp challenge, and we use the 10-core setting. Then Yelp provides 1 million ratings from 45 thousand users on 45 thousand items. MovieLens-20M dataset provides 20 million ratings from 138 thousand users on 27 thousand movies. 
Since our framework focuses on the top-N recommendations, we employed two widely ranking metrics for model evaluation: HR and NDCG~\cite{sun2018attentive}. 
% HR@N calculates the number of predicted top-N items that users are interested in. NDCG@N normalizes the correlation and position of the hit items as an evaluation score~\cite{sun2018attentive}. 
For each user, we select all unrated items as candidates, and mix them with the records in the validation and test sets to select the top-N results.

\begin{table*}[h]
%   \vspace{-0.1cm}
  \footnotesize
  \centering
  \caption{Performance of using different observed set size $K$ and unobserved set size $L$ on Amazon Books. }\label{tab:amazon_multi_setting} 
  \vspace{-0.3cm}
  \begin{tabular}{|l|l|l|l|l|l|l|l|l|l|l|}
  \hline
  \multicolumn{1}{|c|}{\multirow{2}{*}{Models}} & \multicolumn{2}{c|}{N=10} & \multicolumn{2}{c|}{N=20} & \multicolumn{2}{c|}{N=30} & \multicolumn{2}{c|}{N=40} & \multicolumn{2}{c|}{N=50} \\ \cline{2-11} 
  \multicolumn{1}{|c|}{} & HR & NDCG & HR & NDCG & HR & NDCG & HR & NDCG & HR & NDCG \\ \hline
  
  BPR & 0.01851 & 0.01710 & 0.02853 & 0.02169 & 0.03821 & 0.02564 & 0.04737 & 0.02911 & 0.05556 & 0.03205 \\ \hline 
  % BPR-fix(w/o) L2 & 0.01999 & 0.01835 & 0.03128 & 0.02346 & 0.04150 & 0.02762 & 0.05110 & 0.03124 & 0.05973 & 0.03435 \\ \hline
  Set2setRank\_BPR(L=2,K=5) & 0.02084 & 0.01900 & 0.03221 & 0.02417& 0.04276 & 0.02846 & 0.05288 & 0.03228 & 0.06189 & 0.03550 \\ \hline  
  Set2setRank\_BPR(L=3,K=5) & \textbf{0.02144} &\textbf{ 0.0197} &\textbf{ 0.03341} & \textbf{0.02517 }& \textbf{0.04448} & \textbf{0.02968} & \textbf{0.05459} & \textbf{0.03350} & \textbf{0.06378} & \textbf{0.03679} \\ \hline 
  Set2setRank\_BPR(L=4,K=5) & 0.02095 & 0.01904 & 0.03256 & 0.02432 & 0.04366 & 0.02883 & 0.05378 & 0.03265 & 0.06277 & 0.03588 \\ \hline 
  Set2setRank\_BPR(L=5,K=5) & 0.02094 & 0.01893 & 0.03187 & 0.02390 & 0.04209 & 0.02805 & 0.05151 & 0.03160 & 0.06029 & 0.03474 \\ \hline 
  \hline
  Set2setRank\_BPR(L=2,K=10) & 0.02061 & 0.0190 & 0.03224 & 0.02429 & 0.04311 & 0.02873 & 0.05291 & 0.03244 & 0.06182 & 0.03564 \\ \hline  
  Set2setRank\_BPR(L=2,K=20) & \textbf{0.02174} & \textbf{0.01998} & \textbf{0.03413} &\textbf{ 0.02558 }& 0.04566 & 0.03026 & 0.05607 & 0.03420 & 0.06585 & 0.03771 \\ \hline 
  Set2setRank\_BPR(L=2,K=50) & 0.02160 & 0.01987 & 0.03404 & 0.02554 & \textbf{0.04567} & \textbf{0.03027} & \textbf{0.05636} & \textbf{0.03431 }& \textbf{0.06627} & \textbf{0.03785} \\ \hline 
  Set2setRank\_BPR(L=2,K=100) & 0.02166 & 0.01983 & 0.03394 & 0.02541 & 0.04549 & 0.03009 & 0.05595 & 0.03405 & 0.06550 & 0.03745 \\ \hline  
  % BPR-pair(L=2,K=1000) & 0. & 0. & 0. & 0. & 0. & 0. & 0. & 0. & 0. & 0. \\ \hline 
  % BPR-multi & 0.3067 & 0.3197 & 0.3258 & 0.3176 & 0.3567 & 0.3255 & 0.3869 & 0.3355 & 0.3144 & 0.3445 \\ \hline   
  \end{tabular}
  \end{table*}
  
\begin{table*}[h] 
%   \vspace{-0.2cm}
  \footnotesize
  \centering
  \caption{Performance of using different observed set size $K$ and unobserved set size $L$  on MovieLens-20M. }\label{tab:movie_multi_setting} 
  \vspace{-0.3cm}
  \begin{tabular}{|l|l|l|l|l|l|l|l|l|l|l|}
  \hline
  \multicolumn{1}{|c|}{\multirow{2}{*}{Models}} & \multicolumn{2}{c|}{N=10} & \multicolumn{2}{c|}{N=20} & \multicolumn{2}{c|}{N=30} & \multicolumn{2}{c|}{N=40} & \multicolumn{2}{c|}{N=50} \\ \cline{2-11} 
  \multicolumn{1}{|c|}{} & HR & NDCG & HR & NDCG & HR & NDCG & HR & NDCG & HR & NDCG \\ \hline
  BPR & 0.2380 & 0.2504 & 0.2512 & 0.2464 & 0.2767 & 0.2522 & 0.3012 & 0.2593 & 0.3240 & 0.2668 \\ \hline  
  % BPR-fix(w/o) & 0.2899 & 0.3047 & 0.3046 & 0.3005 & 0.3333 & 0.3077 & 0.3623 & 0.3168 & 0.3886 & 0.3257 \\ \hline 
  Set2setRank\_BPR(L=2,K=5) & 0.3072 & 0.3234 & 0.3245 & 0.3195 & 0.3539 & 0.3266 & 0.3838 & 0.3360 & 0.4112 & 0.3453 \\ \hline 
  Set2setRank\_BPR(L=3,K=5) & \textbf{0.3071} & \textbf{0.3204 }& \textbf{0.3235 }& \textbf{0.3169} & \textbf{0.3550} & \textbf{0.3248} & \textbf{0.3857} & \textbf{0.3344} & \textbf{0.4141} & \textbf{0.3439} \\ \hline 
  Set2setRank\_BPR(L=4,K=5) & 0.3052 & 0.3067 & 0.3261 & 0.3077 & 0.3584 & 0.3168 & 0.3899 & 0.3270 & 0.4188 & 0.3369 \\ \hline 
  Set2setRank\_BPR(L=5,K=5) & 0.2902 & 0.2844 & 0.3175 & 0.2914 & 0.3515 & 0.3022 & 0.3835 & 0.3131 & 0.4130 & 0.3235 \\ \hline 
  \hline
  Set2setRank\_BPR(L=2,K=10) & 0.3101 & 0.3262 & 0.3239 & 0.3205 & 0.3550 & 0.3281 & 0.3860 & 0.3378 & 0.4142 & 0.3473 \\ \hline
  Set2setRank\_BPR(L=2,K=20) & 0.3110 & 0.3269 & 0.3280 & 0.3230 & 0.3585 & 0.3305 & 0.3882 & 0.3398 & 0.4151 & 0.3490 \\ \hline  
  Set2setRank\_BPR(L=2,K=50) & \textbf{0.3321} & \textbf{0.3451} & \textbf{0.3559} & \textbf{0.3451} & \textbf{0.3901} & \textbf{0.3548} & \textbf{0.4222} & \textbf{0.3656} & \textbf{0.4517} & \textbf{0.3761} \\ \hline 
  Set2setRank\_BPR(L=2,K=100) & 0.3258 & 0.3360 & 0.3531 & 0.3391 & 0.3903 & 0.3508 & 0.4245 & 0.3629 & 0.4554 & 0.3742 \\ \hline
  % BPR-pair(L=2,K=1000) & 0. & 0. & 0. & 0. & 0. & 0. & 0. & 0. & 0. & 0. \\ \hline 
  % BPR-multi & 0.3067 & 0.3197 & 0.3258 & 0.3176 & 0.3567 & 0.3255 & 0.3869 & 0.3355 & 0.3144 & 0.3445 \\ \hline   
  \end{tabular}
%   \vspace{-0.4cm}
  \end{table*}
  % \end{small}

\textbf{Baseline.}
As mentioned before, our proposed framework is model-agnostic and can be applied to CF-based models. Thus, we adopt two typical models as the base CF model, including the classical matrix factorization model~(BPR)~\cite{rendle2012bpr} and a state-of-the-art graph-based recommendation model~(LR-GCCF)~\cite{chen2020revisiting}. 
For the sake of convenience, we use Set2setRank\_BPR and Set2setRank\_GCN to distinguish different base rating prediction models of our proposed framework. Besides, we use Set2setRank(A)\_BPR and Set2setRank(A)\\\_GCN to denote the adaptive Set2setRank ranking approach that is proposed in Section~\ref{subsec: aset2setrank}.          
In order to better verify the effectiveness of our proposed framework, we select the following baselines:
% \vspace{-0.1cm}
\begin{itemize}
    \item BPR~\cite{rendle2012bpr}: BPR is a widely used pairwise method.% for implicit feedback.
    
    \item LR-GCCF~\cite{chen2020revisiting}: LR-GCCF is a state-of-the-art graph-based recommendation model, which is easier to tune and shows better performance than BRP. We use the pairwise ranking loss as BPR as the optimization goal of LR-CGGF.
    
    \item CLiMF~\cite{shi2012climf}: CLiMF is a typical listwise method for implicit feedback, which optimizes the lower bound of the MRR.% smoothed Mean Reciprocal Rank~(MRR).
    
    \item Deep-SetRank~\cite{wang2020setrank}: Deep-SetRank is a state-of-the-art listwise learning model that achieves better performance than most permutation based ranking models. This method maximizes top-1 permutation probability to guarantee that each user prefers an observed item to multiple unobserved items.% for implicit feedback.%This method is a newly proposed algorithm that assumes that each user prefers the observed item to multiple unobserved items for implicit feedback.%This exploits a new independence assumption for implicit feedback, which assumes that each user prefers the observed item to multiple unobserved items.
    
    \item Pop-Sampling: Instead of the randomly sampling negative items for BPR, Pop-Sampling samples unobserved items based on the popularity of each item in the training data~\cite{rendle2014improving}. 
    
    \item AoBPR~\cite{rendle2014improving}: AoBPR proposed an adaptive and context-dependent sampling distribution to build a non-uniform item sampler. % Instead of uniform negative sampling in BPR,
    
\end{itemize}
% \vspace{-0.1cm}
% We choose these baselines, they are state-of-the-art models for each kind of implicit feedback based recommendation, including pairwise model~(BPR), permutation based listwise model~(Deep-SetRank) and ranking metric optimization based listwise model~(CLiMF), and negative sampling models~(Pop-Sampling and AoBPR).

\textbf{Parameter Settings.}
In this part, we list the parameter settings of Set2setRank. 
The number~$K$ of sampled unobserved set size influences all ranking based models, for fair comparison, we set the number of unobserved item to 5 for all models , including our proposed model. We set~$L=2$ in Set2setRank\_BPR and Set2setRank\_GCN. 
To evaluate the model performance more comprehensively, we select the top-N values in the range of $N=\{10,20,30,40,50\}$. The parameter~$\beta$ is~$\beta=0.5$ for Set2setRank and ~$\beta=0.2$ for Set2setRank(A). The parameter~$\lambda$ is~$\lambda=1$.
We test parameters using the MindSpore~\cite{mindspore} and other tools. We implement the proposed framework using the MindSpore~\cite{mindspore} and other open-source Python libraries.

\begin{table*}[h]
% \vspace{-0.1cm}
  \footnotesize
  \centering
  \caption{Performance under different set to set comparisons, with  ``easy'' means  using the ``easy'' observed sample~( Eq.\eqref{eq:easy_summary}). }\label{tab:add_easy} 
  \vspace{-0.3cm}
  \begin{tabular}{|l|l|l|l|l|l|l|l|l|l|l|l|}
  \hline
  \multirow{2}{*}{Dataset} & \multicolumn{1}{c|}{\multirow{2}{*}{Models}} & \multicolumn{2}{c|}{N=10} & \multicolumn{2}{c|}{N=20} & \multicolumn{2}{c|}{N=30} & \multicolumn{2}{c|}{N=40} & \multicolumn{2}{c|}{N=50} \\ \cline{3-12} 
      & \multicolumn{1}{c|}{} & HR & NDCG & HR & NDCG & HR & NDCG & HR & NDCG & HR & NDCG \\ \hline
    \multirow{4}{*}{Amazon Books} & Set2setRank\_BPR(L=3,K=5)easy & 0.02119 & 0.01919 & 0.03334 & 0.02476 & 0.04458 & 0.02934 & 0.05476 & 0.03319 & 0.06385 & 0.03645 \\ \cline{2-12} 
    & Set2setRank\_BPR(L=3,K=5) & 0.02144 & 0.0197 & 0.03341 & 0.02517 & 0.04448 & 0.02968 & 0.05459 & 0.03350 & 0.06378 & 0.03679 \\ \cline{2-12}  
    & Set2setRank\_BPR(L=4,K=5)easy & 0.01946 & 0.01780 & 0.03072 & 0.02291 & 0.04062 & 0.02693 & 0.04972 & 0.03037 & 0.05811 & 0.03336 \\ \cline{2-12} 
      & Set2setRank\_BPR(L=4,K=5) & 0.02095 & 0.01904 & 0.03256 & 0.02432 & 0.04366 & 0.02883 & 0.05378 & 0.03265 & 0.06277 & 0.03588 \\ \hline
  \multirow{4}{*}{MovieLens-20M} & Set2setRank\_BPR(L=3,K=5)easy & 0.3075 & 0.3196 & 0.3265 & 0.3177 & 0.3591 & 0.3263 & 0.3903 & 0.3362 & 0.4183 & 0.3456   \\ \cline{2-12} 
  & Set2setRank\_BPR(L=3,K=5) & 0.3071 & 0.3204 & 0.3235 & 0.3169 & 0.3550 & 0.3248 & 0.3857 & 0.3344 & 0.4141 & 0.3439 \\ \cline{2-12}
  & Set2setRank\_BPR(L=4,K=5)easy & 0.2994 & 0.2976 & 0.3267 & 0.3028 & 0.3615 & 0.3134 & 0.3933 & 0.3241 & 0.4220 & 0.3342 \\ \cline{2-12}
      & Set2setRank\_BPR(L=4,K=5) & 0.3052 & 0.3067 & 0.3261 & 0.3077 & 0.3584 & 0.3168 & 0.3899 & 0.3270 & 0.4188 & 0.3369  \\  \hline
  \end{tabular}
\end{table*}

\begin{table*}[h] 
% \vspace{-0.1cm}
  \footnotesize
  \centering
  \caption{Performance of the two modules of Set2setRank on AmazonBooks: item to set comparison and set to set comparison.}\label{tab:abla_amazon}  
  \vspace{-0.3cm}
  \begin{tabular}{|l|l|l|l|l|l|l|l|l|l|l|l|l|}
  \hline
  \multicolumn{1}{|c|}{\multirow{2}{*}{Models}} & \multirow{2}{*}{\begin{tabular}[c]{@{}l@{}}Item to set\\ comparison\end{tabular}} & \multirow{2}{*}{\begin{tabular}[c]{@{}l@{}}Set to set\\ comparison\end{tabular}} & \multicolumn{2}{c|}{N=10} & \multicolumn{2}{c|}{N=20} & \multicolumn{2}{c|}{N=30} & \multicolumn{2}{c|}{N=40} & \multicolumn{2}{c|}{N=50} \\ \cline{4-13} 
  \multicolumn{1}{|c|}{} &  &  & HR & NDCG & HR & NDCG & HR & NDCG & HR & NDCG & HR & NDCG \\ \hline
  BPR & / &/  & 0.01851 & 0.01710 & 0.02853 & 0.02169 & 0.03821 & 0.02564 & 0.04737 & 0.02911 & 0.05556 & 0.03205 \\ \hline
  Set2setRank\_BPR & \ding{51} & \ding{53} & 0.01999 & 0.01835 & 0.03128 & 0.02346 & 0.04150 & 0.02762 & 0.05110 & 0.03124 & 0.05973 & 0.03435 \\ \hline
  Set2setRank\_BPR & \ding{53} & \ding{51}  & 0.02019 & 0.01872 & 0.02978 & 0.02309 & 0.03946 & 0.02704 & 0.04835 & 0.03042 & 0.05651 & 0.03335 \\ \hline
  Set2setRank\_BPR & \ding{51} & \ding{51} & \textbf{0.02084 }& \textbf{0.01900} & \textbf{0.03221} & \textbf{0.02417}& \textbf{0.04276} & \textbf{0.02846} & \textbf{0.05288} & \textbf{0.03228} & \textbf{0.06189}& \textbf{0.03550} \\ \hline
  LR-GCCF & / & / & 0.02209& 0.02040 & 0.03407 & 0.02583 & 0.04532 & 0.03039 & 0.05532 & 0.03416 & 0.06498 & 0.03761 \\ \hline
  Set2setRank\_GCN & \ding{51} & \ding{53} & 0.02237 & 0.02052 & 0.03459 & 0.02605 & 0.04658 & 0.03092 & 0.05739 & 0.03503 & 0.06770 & 0.03871 \\ \hline
  Set2setRank\_GCN & \ding{53}& \ding{51} & 0.02250 & 0.02076 & 0.03465 & 0.02628 & 0.04603 & 0.03093 & 0.05628 & 0.03480 & 0.06592 & 0.03824 \\ \hline
  Set2setRank\_GCN & \ding{51} & \ding{51} & \textbf{0.02247} & \textbf{0.02079} & \textbf{0.03454} & \textbf{0.02629 }& \textbf{0.04567} & \textbf{0.03089} & \textbf{0.05648} & \textbf{0.03490} & \textbf{0.06597} & \textbf{0.03830} \\ \hline
  \end{tabular}
\end{table*}

\begin{table*}[h]
\vspace{-0.2cm}
  \footnotesize
  \centering
  \caption{Performance of the two modules of Set2setRank on MovieLens-20M: item to set comparison and set to set comparison. }\label{tab:abla_movie}
  \vspace{-0.3cm}
  \begin{tabular}{|l|l|l|l|l|l|l|l|l|l|l|l|l|}
  \hline
  \multicolumn{1}{|c|}{\multirow{2}{*}{Models}} & \multirow{2}{*}{\begin{tabular}[c]{@{}l@{}}Item to set\\ comparison\end{tabular}} & \multirow{2}{*}{\begin{tabular}[c]{@{}l@{}}Set to set\\ comparison\end{tabular}} & \multicolumn{2}{c|}{N=10} & \multicolumn{2}{c|}{N=20} & \multicolumn{2}{c|}{N=30} & \multicolumn{2}{c|}{N=40} & \multicolumn{2}{c|}{N=50} \\ \cline{4-13} 
  \multicolumn{1}{|c|}{} &  &  & HR & NDCG & HR & NDCG & HR & NDCG & HR & NDCG & HR & NDCG \\ \hline
  BPR & / &/  & 0.2380 & 0.2504 & 0.2512 & 0.2464 & 0.2767 & 0.2522 & 0.3012 & 0.2593 & 0.3240 & 0.2668 \\ \hline 
  Set2setRank\_BPR & \ding{51} & \ding{53} & 0.2899 & 0.3047 & 0.3046 & 0.3005 & 0.3333 & 0.3077 & 0.3623 & 0.3168 & 0.3886 & 0.3257 \\ \hline
  Set2setRank\_BPR & \ding{53} & \ding{51} & 0.2869 & 0.3057 & 0.2984 & 0.2979 & 0.3246 & 0.3033 & 0.3510 & 0.3109 & 0.3761 & 0.3190 \\ \hline%e70
  Set2setRank\_BPR & \ding{51} & \ding{51} & \textbf{0.3072} & \textbf{0.3234} & \textbf{0.3245} & \textbf{0.3195} & \textbf{0.3539} & \textbf{0.3266 }& \textbf{0.3838} & \textbf{0.3360} & \textbf{0.4112} & \textbf{0.3453} \\ \hline
  LR-GCCF & / & /  & 0.2672 & 0.2764 & 0.2832 & 0.2752 & 0.3125 & 0.2837 & 0.3426 & 0.2929 & 0.3715 & 0.3042 \\ \hline 
  Set2setRank\_GCN & \ding{51} & \ding{53} & 0.2671 & 0.2777 & 0.2832 & 0.2760 & 0.3125 & 0.2843 & 0.3422 & 0.2942 & 0.3700 & 0.3040 \\ \hline 
  Set2setRank\_GCN & \ding{53}& \ding{51} & 0.2650 & 0.2793 & 0.2773 & 0.2746 & 0.3042 & 0.2811 & 0.3318 & 0.2896 & 0.3583 & 0.2984 \\ \hline%s5e45
  Set2setRank\_GCN& \ding{51} & \ding{51} & \textbf{0.2689} & \textbf{0.2801} & \textbf{0.2832} & \textbf{0.2771} & \textbf{0.3114} & \textbf{0.2847} & \textbf{0.3403} & \textbf{0.2942} & \textbf{0.3675} & \textbf{0.3036} \\ \hline
  \end{tabular}
\end{table*}

\subsection{Experimental Results and Analyses}

\subsubsection{Overall Performance Comparison}

Tables~\ref{tab:resall} reports the overall experimental results on different datasets with different evaluation metrics. 
We can obtain the following observations:
\begin{itemize}
    \item Our proposed Set2setRank achieves the best performance across all datasets with different evaluation metrics. Specifically, Set2setRank\_BPR(A) outperforms the best baselines by an average relative boost of $9\%, 11\%$ and $4\%$ on Amazon Books MovieLens-20M, and Yelp, respectively. 
    Set2setRank\\\_GCN(A) also outperforms all the baselines. The adaptive setting usually performs better than the counterpart without adaptive sampling. These observations demonstrate that our proposed two-level set comparison can not only fully exploit implicit feedback to model user preference, but also provide accurate recommendation results. Moreover, based on Set2setRank, the improvement of BPR is better than LR-GCCF. The reason is that LR-GCCF uses more characteristics of implicit feedback than BPR, which can further prove the effectiveness of our proposed Set2setRank framework on implicit feedback utilization. 
    
    \item Set2setRank\_GCN (Set2setRank\_BPR) also achieves promising performance on all datasets. However, the corresponding improvements are not so obvious than Set2setRank\_GCN (Set2setRank\_BPR), demonstrating that our newly designed adaptive sampling technique can further improve the model performance. 
    Moreover, we obtain that the improvement is not much. 
    We speculate one possible reason is that the length of unobserved samples set and observed samples set in this experiment is small, which limits the ability of our proposed adaptive sampling technique. 
    
    \item Among all baselines, CLiMF does not achieve satisfactory performance on all datasets. This is because CLiMF focuses on the first relevant item of recommendation lists. This situation also appears in~\cite{yu2020collaborative,liang2018top}. Meanwhile, Deep-SetRank outperforms the pairwise learning and listwise learning method, this is also consistent with the results in Deep-SetRank~\cite{wang2020setrank}. 
\end{itemize}
% \vspace{-0.3cm}

\subsubsection{Influence of Set Size.} 
Since we propose to compare the observed samples and unobserved samples at set level, the sizes of these two sets play an important role. In order to explore the influence of different negative set size $K$ and positive set size $L$, we make additional experiments on Amazon Books and MovieLens-20M datasets to verify the model performance. Moreover, we select Set2setRank\_BPR to avoid the influence of our proposed adapted sampling technique. 
The results are reported in Table~\ref{tab:amazon_multi_setting} and~\ref{tab:movie_multi_setting}. 
From these two tables, we can observe that with the increase of~$K$ and~$L$, the results first increase and then decrease. 
The best setting is $L=3, K=50$ for Amazon Book dataset, and $L=3, K=50$ for MovieLens-20M dataset. 
On the one hand, with the increase of~$K$ and~$L$, the model can compare more observed samples and unobserved samples, which is very helpful for hidden structure information exploration and implicit feedback utilization. 
On the other hand, since the size of unobserved items is much larger than observed items, continually increasing $K$ and~$L$ will lead to the inconsistency of sampled observed set and sampled unobserved set, which will result  model performance drop. 
This observation inspires us to carefully determine the value of $K$ and~$L$ for the best performance of our proposed framework.

\subsubsection{Set to Set Comparison Strategy}
In Section~\ref{s:set2set_compare}, we have mentioned that the set to set comparison in our proposed two-level comparison has two strategies: 1) in Eq.\eqref{eq:pos_summary}, considering the relationships of all observed pairs as the summary of the observed set.
2) in Eq.\eqref{eq:easy_summary}, selecting the ``easy'' observed sample with the consideration of the observed set. In this part, we compare the performance of these two set to set comparison strategies.
We employ Set2setRank\_BPR as the base model and make additional experiments to verify the effectiveness of each strategy. 
Table~\ref{tab:add_easy} summaries the corresponding results on Amazon Books and MovieLens-20M datasets. 
From the results, we can conclude that both strategies achieve relatively high performance on two datsets with different evaluation metrics, in which summarizing all observed pairs achieves better performance.  
We speculate a possible reason is that selecting the ``hard'' unobserved sample already fully exploits the observed set and is also an effective way to improve performance, as shown by the negative sampling model~\cite{rendle2014improving,lian2020personalized}. If we further select ``easy'' positive item, this positive item information has already been calculated from item to set level comparison, and set to set comparison degenerates to increase weights of ``easy'' positive item. In contrast, the  summary information of the observed set is not well exploited. Therefore, considering the relationships of all observed pairs in Eq.\eqref{eq:pos_summary} shows the best performance. To this end, we select the strategy in Eq.\eqref{eq:pos_summary} to implement our proposed framework.

\subsubsection{Ablation Study}
In this part, we investigate the effectiveness of each component:
item to set comparison~(Eq.\eqref{eq:npair_loss}) and set to set comparison~(Eq.\eqref{eq:set_loss}) of our proposed Set2set framework.
The results are illustrated in Tables~\ref{tab:abla_amazon} and~\ref{tab:abla_movie}.
From the two tables, we can obtain the following observations. First,
each single component~(item to set comparison or set to set comparison) can still help model achieve comparable performance, indicating the usefulness of our proposed comparison at set level. Moreover, with single component, the model achieves similar performance, indicating that both of them are very important for user preference modeling and model performance improvement. Besides, compared with the performance of models with single component, models with both of them~(entire Set2setRank framework) have better performance on all datasets, demonstrating that necessity of both components in our proposed two-level comparison. As these two components consider different self-supervised information for implicit feedback, combining them together reaches the best performance.

% \section{Conclusion and Future Work}
\section{Conclusion}
In this paper, we argued that current ranking models for CF are still far from satisfactory due to the sparsity of the observed user-item interactions for each user. 
To this end, we proposed to compare the observed items and unobserved items at set level to fully explore the structure information hidden in the implicit feedback.Specifically, we first constructed the sampled observed set and sampled unobserved set. Then, we proposed a Set2setRank framework for the utilization of implicit feedback. 
Set2setRank mainly consisted of two parts: item to set comparison to encourage unobserved samples are ranked lower than the observed set, and set to set comparison to fully explore the limited observed set to be far away from the hard negative sample from the unobserved set.
Moreover, we designed an adaptive technique to further improve the performance of Set2setRank framework. Extensive experiments on three real-world datasets clearly showed the advantages of our proposed framework.

\begin{acks}
This work was supported in part by grants from the National Natural Science Foundation of China (Grant No. 61972125, U1936219, 61932009, 61732008, 62006066), the Young Elite Scientists Sponsorship Program by CAST and ISZS, CAAI-Huawei MindSpore Open Fund, and the Fundamental Research Funds for the Central Universities, HFUT.
\end{acks}

\newpage
\bibliographystyle{ACM-Reference-Format}
\bibliography{paperlist}


\begin{thebibliography}{55}

%%% the Web address.

\ifx \showCODEN    \undefined \def \showCODEN     #1{\unskip}     \fi
\ifx \showDOI      \undefined \def \showDOI       #1{#1}\fi
\ifx \showISBNx    \undefined \def \showISBNx     #1{\unskip}     \fi
\ifx \showISBNxiii \undefined \def \showISBNxiii  #1{\unskip}     \fi
\ifx \showISSN     \undefined \def \showISSN      #1{\unskip}     \fi
\ifx \showLCCN     \undefined \def \showLCCN      #1{\unskip}     \fi
\ifx \shownote     \undefined \def \shownote      #1{#1}          \fi
\ifx \showarticletitle \undefined \def \showarticletitle #1{#1}   \fi
\ifx \showURL      \undefined \def \showURL       {\relax}        \fi
% invisible to TeX
\providecommand\bibfield[2]{#2}
\providecommand\bibinfo[2]{#2}
\providecommand\natexlab[1]{#1}
\providecommand\showeprint[2][]{arXiv:#2}

\bibitem[\protect\citeauthoryear{Balakrishnan and Chopra}{Balakrishnan and
  Chopra}{2012}]%
        {balakrishnan2012collaborative}
\bibfield{author}{\bibinfo{person}{Suhrid Balakrishnan} {and}
  \bibinfo{person}{Sumit Chopra}.} \bibinfo{year}{2012}\natexlab{}.
\newblock \showarticletitle{Collaborative ranking}. In
  \bibinfo{booktitle}{\emph{WSDM}}. \bibinfo{pages}{143--152}.
\newblock


\bibitem[\protect\citeauthoryear{Cao, Qin, Liu, Tsai, and Li}{Cao
  et~al\mbox{.}}{2007}]%
        {cao2007learning}
\bibfield{author}{\bibinfo{person}{Zhe Cao}, \bibinfo{person}{Tao Qin},
  \bibinfo{person}{Tie-Yan Liu}, \bibinfo{person}{Ming-Feng Tsai}, {and}
  \bibinfo{person}{Hang Li}.} \bibinfo{year}{2007}\natexlab{}.
\newblock \showarticletitle{Learning to rank: from pairwise approach to
  listwise approach}. In \bibinfo{booktitle}{\emph{ICML}}.
  \bibinfo{pages}{129--136}.
\newblock


\bibitem[\protect\citeauthoryear{Chae, Kim, Chau, and Kim}{Chae
  et~al\mbox{.}}{2020}]%
        {chae2020ar}
\bibfield{author}{\bibinfo{person}{Dong-Kyu Chae}, \bibinfo{person}{Jihoo Kim},
  \bibinfo{person}{Duen~Horng Chau}, {and} \bibinfo{person}{Sang-Wook Kim}.}
  \bibinfo{year}{2020}\natexlab{}.
\newblock \showarticletitle{AR-CF: Augmenting Virtual Users and Items in
  Collaborative Filtering for Addressing Cold-Start Problems}. In
  \bibinfo{booktitle}{\emph{SIGIR}}. \bibinfo{pages}{1251--1260}.
\newblock


\bibitem[\protect\citeauthoryear{Chen, Zhang, Ma, Liu, and Ma}{Chen
  et~al\mbox{.}}{2020c}]%
        {chen2020jointly}
\bibfield{author}{\bibinfo{person}{Chong Chen}, \bibinfo{person}{Min Zhang},
  \bibinfo{person}{Weizhi Ma}, \bibinfo{person}{Yiqun Liu}, {and}
  \bibinfo{person}{Shaoping Ma}.} \bibinfo{year}{2020}\natexlab{c}.
\newblock \showarticletitle{Jointly non-sampling learning for knowledge graph
  enhanced recommendation}. In \bibinfo{booktitle}{\emph{SIGIR}}.
  \bibinfo{pages}{189--198}.
\newblock


\bibitem[\protect\citeauthoryear{Chen, Wu, Hong, Zhang, and Wang}{Chen
  et~al\mbox{.}}{2020a}]%
        {chen2020revisiting}
\bibfield{author}{\bibinfo{person}{Lei Chen}, \bibinfo{person}{Le Wu},
  \bibinfo{person}{Richang Hong}, \bibinfo{person}{Kun Zhang}, {and}
  \bibinfo{person}{Meng Wang}.} \bibinfo{year}{2020}\natexlab{a}.
\newblock \showarticletitle{Revisiting Graph Based Collaborative Filtering: A
  Linear Residual Graph Convolutional Network Approach}. In
  \bibinfo{booktitle}{\emph{AAAI}}, Vol.~\bibinfo{volume}{34}.
  \bibinfo{pages}{27--34}.
\newblock


\bibitem[\protect\citeauthoryear{Chen, Xiao, Li, Ye, Sun, and Deng}{Chen
  et~al\mbox{.}}{2020b}]%
        {chen2020esam}
\bibfield{author}{\bibinfo{person}{Zhihong Chen}, \bibinfo{person}{Rong Xiao},
  \bibinfo{person}{Chenliang Li}, \bibinfo{person}{Gangfeng Ye},
  \bibinfo{person}{Haochuan Sun}, {and} \bibinfo{person}{Hongbo Deng}.}
  \bibinfo{year}{2020}\natexlab{b}.
\newblock \showarticletitle{Esam: Discriminative domain adaptation with
  non-displayed items to improve long-tail performance}. In
  \bibinfo{booktitle}{\emph{SIGIR}}. \bibinfo{pages}{579--588}.
\newblock


\bibitem[\protect\citeauthoryear{Cui, Che, Liu, Qin, Yang, Wang, and Hu}{Cui
  et~al\mbox{.}}{2020}]%
        {cui2019pre}
\bibfield{author}{\bibinfo{person}{Yiming Cui}, \bibinfo{person}{Wanxiang Che},
  \bibinfo{person}{Ting Liu}, \bibinfo{person}{Bing Qin},
  \bibinfo{person}{Ziqing Yang}, \bibinfo{person}{Shijin Wang}, {and}
  \bibinfo{person}{Guoping Hu}.} \bibinfo{year}{2020}\natexlab{}.
\newblock \showarticletitle{Pre-training with whole word masking for chinese
  bert}. In \bibinfo{booktitle}{\emph{EMNLP}}. \bibinfo{pages}{657--668}.
\newblock


\bibitem[\protect\citeauthoryear{Devlin, Chang, Lee, and Toutanova}{Devlin
  et~al\mbox{.}}{2019}]%
        {devlin2018bert}
\bibfield{author}{\bibinfo{person}{Jacob Devlin}, \bibinfo{person}{Ming-Wei
  Chang}, \bibinfo{person}{Kenton Lee}, {and} \bibinfo{person}{Kristina
  Toutanova}.} \bibinfo{year}{2019}\natexlab{}.
\newblock \showarticletitle{BERT: Pre-training of deep bidirectional
  transformers for language understanding}. In
  \bibinfo{booktitle}{\emph{NAACL}}. \bibinfo{pages}{4171--4186}.
\newblock


\bibitem[\protect\citeauthoryear{Ding, Feng, He, Yu, Li, and Jin}{Ding
  et~al\mbox{.}}{2018}]%
        {ding2018improved}
\bibfield{author}{\bibinfo{person}{Jingtao Ding}, \bibinfo{person}{Fuli Feng},
  \bibinfo{person}{Xiangnan He}, \bibinfo{person}{Guanghui Yu},
  \bibinfo{person}{Yong Li}, {and} \bibinfo{person}{Depeng Jin}.}
  \bibinfo{year}{2018}\natexlab{}.
\newblock \showarticletitle{An improved sampler for bayesian personalized
  ranking by leveraging view data}. In \bibinfo{booktitle}{\emph{WWW}}.
  \bibinfo{pages}{13--14}.
\newblock


\bibitem[\protect\citeauthoryear{Gantner, Drumond, Freudenthaler, and
  Schmidt-Thieme}{Gantner et~al\mbox{.}}{2012}]%
        {gantner2012personalized}
\bibfield{author}{\bibinfo{person}{Zeno Gantner}, \bibinfo{person}{Lucas
  Drumond}, \bibinfo{person}{Christoph Freudenthaler}, {and}
  \bibinfo{person}{Lars Schmidt-Thieme}.} \bibinfo{year}{2012}\natexlab{}.
\newblock \showarticletitle{Personalized ranking for non-uniformly sampled
  items}. In \bibinfo{booktitle}{\emph{SIGKDD}}. \bibinfo{pages}{231--247}.
\newblock


\bibitem[\protect\citeauthoryear{Grbovic and Cheng}{Grbovic and Cheng}{2018}]%
        {grbovic2018real}
\bibfield{author}{\bibinfo{person}{Mihajlo Grbovic} {and}
  \bibinfo{person}{Haibin Cheng}.} \bibinfo{year}{2018}\natexlab{}.
\newblock \showarticletitle{Real-time personalization using embeddings for
  search ranking at airbnb}. In \bibinfo{booktitle}{\emph{SIGKDD}}.
  \bibinfo{pages}{311--320}.
\newblock


\bibitem[\protect\citeauthoryear{He, Deng, Wang, Li, Zhang, and Wang}{He
  et~al\mbox{.}}{2020}]%
        {he2020lightgcn}
\bibfield{author}{\bibinfo{person}{Xiangnan He}, \bibinfo{person}{Kuan Deng},
  \bibinfo{person}{Xiang Wang}, \bibinfo{person}{Yan Li},
  \bibinfo{person}{Yongdong Zhang}, {and} \bibinfo{person}{Meng Wang}.}
  \bibinfo{year}{2020}\natexlab{}.
\newblock \showarticletitle{Lightgcn: Simplifying and powering graph
  convolution network for recommendation}. In
  \bibinfo{booktitle}{\emph{SIGIR}}. \bibinfo{pages}{639--648}.
\newblock


\bibitem[\protect\citeauthoryear{Hsieh, Yang, Cui, Lin, Belongie, and
  Estrin}{Hsieh et~al\mbox{.}}{2017}]%
        {hsieh2017collaborative}
\bibfield{author}{\bibinfo{person}{Cheng-Kang Hsieh}, \bibinfo{person}{Longqi
  Yang}, \bibinfo{person}{Yin Cui}, \bibinfo{person}{Tsung-Yi Lin},
  \bibinfo{person}{Serge Belongie}, {and} \bibinfo{person}{Deborah Estrin}.}
  \bibinfo{year}{2017}\natexlab{}.
\newblock \showarticletitle{Collaborative metric learning}. In
  \bibinfo{booktitle}{\emph{WWW}}. \bibinfo{pages}{193--201}.
\newblock


\bibitem[\protect\citeauthoryear{Huang, Wang, Liu, Ma, Chen, and
  Veijalainen}{Huang et~al\mbox{.}}{2015}]%
        {huang2015listwise}
\bibfield{author}{\bibinfo{person}{Shanshan Huang}, \bibinfo{person}{Shuaiqiang
  Wang}, \bibinfo{person}{Tie-Yan Liu}, \bibinfo{person}{Jun Ma},
  \bibinfo{person}{Zhumin Chen}, {and} \bibinfo{person}{Jari Veijalainen}.}
  \bibinfo{year}{2015}\natexlab{}.
\newblock \showarticletitle{Listwise collaborative filtering}. In
  \bibinfo{booktitle}{\emph{SIGIR}}. \bibinfo{pages}{343--352}.
\newblock


\bibitem[\protect\citeauthoryear{Jamali and Ester}{Jamali and Ester}{2010}]%
        {jamali2010matrix}
\bibfield{author}{\bibinfo{person}{Mohsen Jamali} {and} \bibinfo{person}{Martin
  Ester}.} \bibinfo{year}{2010}\natexlab{}.
\newblock \showarticletitle{A matrix factorization technique with trust
  propagation for recommendation in social networks}. In
  \bibinfo{booktitle}{\emph{RecSys}}. \bibinfo{pages}{135--142}.
\newblock


\bibitem[\protect\citeauthoryear{Koren, Bell, and Volinsky}{Koren
  et~al\mbox{.}}{2009}]%
        {koren2009matrix}
\bibfield{author}{\bibinfo{person}{Yehuda Koren}, \bibinfo{person}{Robert
  Bell}, {and} \bibinfo{person}{Chris Volinsky}.}
  \bibinfo{year}{2009}\natexlab{}.
\newblock \showarticletitle{Matrix factorization techniques for recommender
  systems}.
\newblock \bibinfo{journal}{\emph{Computer}} \bibinfo{volume}{42},
  \bibinfo{number}{8} (\bibinfo{year}{2009}), \bibinfo{pages}{30--37}.
\newblock


\bibitem[\protect\citeauthoryear{Lian, Liu, and Chen}{Lian
  et~al\mbox{.}}{2020}]%
        {lian2020personalized}
\bibfield{author}{\bibinfo{person}{Defu Lian}, \bibinfo{person}{Qi Liu}, {and}
  \bibinfo{person}{Enhong Chen}.} \bibinfo{year}{2020}\natexlab{}.
\newblock \showarticletitle{Personalized Ranking with Importance Sampling}. In
  \bibinfo{booktitle}{\emph{WWW}}. \bibinfo{pages}{1093--1103}.
\newblock


\bibitem[\protect\citeauthoryear{Liang, Hu, Dong, and Honavar}{Liang
  et~al\mbox{.}}{2018}]%
        {liang2018top}
\bibfield{author}{\bibinfo{person}{Junjie Liang}, \bibinfo{person}{Jinlong Hu},
  \bibinfo{person}{Shoubin Dong}, {and} \bibinfo{person}{Vasant Honavar}.}
  \bibinfo{year}{2018}\natexlab{}.
\newblock \showarticletitle{Top-N-Rank: A Scalable List-wise Ranking Method for
  Recommender Systems}. In \bibinfo{booktitle}{\emph{Big Data}}.
  \bibinfo{pages}{1052--1058}.
\newblock


\bibitem[\protect\citeauthoryear{Liu, Xiang, Zhao, and Yang}{Liu
  et~al\mbox{.}}{2010}]%
        {liu2010unifying}
\bibfield{author}{\bibinfo{person}{Nathan~N Liu}, \bibinfo{person}{Evan~W
  Xiang}, \bibinfo{person}{Min Zhao}, {and} \bibinfo{person}{Qiang Yang}.}
  \bibinfo{year}{2010}\natexlab{}.
\newblock \showarticletitle{Unifying explicit and implicit feedback for
  collaborative filtering}. In \bibinfo{booktitle}{\emph{ICKM}}.
  \bibinfo{pages}{1445--1448}.
\newblock


\bibitem[\protect\citeauthoryear{Liu, Huang, Yin, Chen, Xiong, Su, and Hu}{Liu
  et~al\mbox{.}}{2019}]%
        {liu2019ekt}
\bibfield{author}{\bibinfo{person}{Qi Liu}, \bibinfo{person}{Zhenya Huang},
  \bibinfo{person}{Yu Yin}, \bibinfo{person}{Enhong Chen}, \bibinfo{person}{Hui
  Xiong}, \bibinfo{person}{Yu Su}, {and} \bibinfo{person}{Guoping Hu}.}
  \bibinfo{year}{2019}\natexlab{}.
\newblock \showarticletitle{Ekt: Exercise-aware knowledge tracing for student
  performance prediction}.
\newblock \bibinfo{journal}{\emph{TKDE}} \bibinfo{volume}{33},
  \bibinfo{number}{1} (\bibinfo{year}{2019}), \bibinfo{pages}{100--115}.
\newblock


\bibitem[\protect\citeauthoryear{Liu, Zhao, Sun, and Miao}{Liu
  et~al\mbox{.}}{2015}]%
        {liu2015boosting}
\bibfield{author}{\bibinfo{person}{Yong Liu}, \bibinfo{person}{Peilin Zhao},
  \bibinfo{person}{Aixin Sun}, {and} \bibinfo{person}{Chunyan Miao}.}
  \bibinfo{year}{2015}\natexlab{}.
\newblock \showarticletitle{A boosting algorithm for item recommendation with
  implicit feedback}. In \bibinfo{booktitle}{\emph{IJCAI}}.
  \bibinfo{pages}{1792--1798}.
\newblock


\bibitem[\protect\citeauthoryear{Lu, Zhang, Ma, Wang, xia, Liu, Lin, and Ma}{Lu
  et~al\mbox{.}}{2019}]%
        {lu2019effects}
\bibfield{author}{\bibinfo{person}{Hongyu Lu}, \bibinfo{person}{Min Zhang},
  \bibinfo{person}{Weizhi Ma}, \bibinfo{person}{Ce Wang}, \bibinfo{person}{Feng
  xia}, \bibinfo{person}{Yiqun Liu}, \bibinfo{person}{Leyu Lin}, {and}
  \bibinfo{person}{Shaoping Ma}.} \bibinfo{year}{2019}\natexlab{}.
\newblock \showarticletitle{Effects of User Negative Experience in Mobile News
  Streaming}. In \bibinfo{booktitle}{\emph{SIGIR}}. \bibinfo{pages}{705--714}.
\newblock


\bibitem[\protect\citeauthoryear{Mindspore}{Mindspore}{2020}]%
        {mindspore}
\bibfield{author}{\bibinfo{person}{Mindspore}.}
  \bibinfo{year}{2020}\natexlab{}.
\newblock \bibinfo{booktitle}{}.
\newblock
\urldef\tempurl%
\url{https://www.mindspore.cn/}
\showURL{%
\tempurl}


\bibitem[\protect\citeauthoryear{Ouyang, Li, Pan, and Ming}{Ouyang
  et~al\mbox{.}}{2019}]%
        {ouyang2019asymmetric}
\bibfield{author}{\bibinfo{person}{Shan Ouyang}, \bibinfo{person}{Lin Li},
  \bibinfo{person}{Weike Pan}, {and} \bibinfo{person}{Zhong Ming}.}
  \bibinfo{year}{2019}\natexlab{}.
\newblock \showarticletitle{Asymmetric Bayesian personalized ranking for
  one-class collaborative filtering}. In \bibinfo{booktitle}{\emph{RecSys}}.
  \bibinfo{pages}{373--377}.
\newblock


\bibitem[\protect\citeauthoryear{Pan, Zhou, Cao, Liu, Lukose, Scholz, and
  Yang}{Pan et~al\mbox{.}}{2008}]%
        {pan2008one}
\bibfield{author}{\bibinfo{person}{Rong Pan}, \bibinfo{person}{Yunhong Zhou},
  \bibinfo{person}{Bin Cao}, \bibinfo{person}{Nathan~N Liu},
  \bibinfo{person}{Rajan Lukose}, \bibinfo{person}{Martin Scholz}, {and}
  \bibinfo{person}{Qiang Yang}.} \bibinfo{year}{2008}\natexlab{}.
\newblock \showarticletitle{One-class collaborative filtering}. In
  \bibinfo{booktitle}{\emph{ICDM}}. \bibinfo{pages}{502--511}.
\newblock


\bibitem[\protect\citeauthoryear{Pan and Chen}{Pan and Chen}{2013a}]%
        {pan2013cofiset}
\bibfield{author}{\bibinfo{person}{Weike Pan} {and} \bibinfo{person}{Li Chen}.}
  \bibinfo{year}{2013}\natexlab{a}.
\newblock \showarticletitle{Cofiset: Collaborative filtering via learning
  pairwise preferences over item-sets}. In \bibinfo{booktitle}{\emph{SIAM}}.
  \bibinfo{pages}{180--188}.
\newblock


\bibitem[\protect\citeauthoryear{Pan and Chen}{Pan and Chen}{2013b}]%
        {pan2013gbpr}
\bibfield{author}{\bibinfo{person}{Weike Pan} {and} \bibinfo{person}{Li Chen}.}
  \bibinfo{year}{2013}\natexlab{b}.
\newblock \showarticletitle{Gbpr: Group preference based bayesian personalized
  ranking for one-class collaborative filtering}. In
  \bibinfo{booktitle}{\emph{IJCAI}}. \bibinfo{pages}{2691--2697}.
\newblock


\bibitem[\protect\citeauthoryear{Pang, Xu, Ai, Lan, Cheng, and Wen}{Pang
  et~al\mbox{.}}{2020}]%
        {pang2020setrank}
\bibfield{author}{\bibinfo{person}{Liang Pang}, \bibinfo{person}{Jun Xu},
  \bibinfo{person}{Qingyao Ai}, \bibinfo{person}{Yanyan Lan},
  \bibinfo{person}{Xueqi Cheng}, {and} \bibinfo{person}{Jirong Wen}.}
  \bibinfo{year}{2020}\natexlab{}.
\newblock \showarticletitle{Setrank: Learning a permutation-invariant ranking
  model for information retrieval}. In \bibinfo{booktitle}{\emph{SIGIR}}.
  \bibinfo{pages}{499--508}.
\newblock


\bibitem[\protect\citeauthoryear{Qi, Wang, Hu, Li, He, and Xu}{Qi
  et~al\mbox{.}}{2019}]%
        {qi2019time}
\bibfield{author}{\bibinfo{person}{Lianyong Qi}, \bibinfo{person}{Ruili Wang},
  \bibinfo{person}{Chunhua Hu}, \bibinfo{person}{Shancang Li},
  \bibinfo{person}{Qiang He}, {and} \bibinfo{person}{Xiaolong Xu}.}
  \bibinfo{year}{2019}\natexlab{}.
\newblock \showarticletitle{Time-aware distributed service recommendation with
  privacy-preservation}.
\newblock \bibinfo{journal}{\emph{Information Sciences}}  \bibinfo{volume}{480}
  (\bibinfo{year}{2019}), \bibinfo{pages}{354--364}.
\newblock


\bibitem[\protect\citeauthoryear{Rendle and Freudenthaler}{Rendle and
  Freudenthaler}{2014}]%
        {rendle2014improving}
\bibfield{author}{\bibinfo{person}{Steffen Rendle} {and}
  \bibinfo{person}{Christoph Freudenthaler}.} \bibinfo{year}{2014}\natexlab{}.
\newblock \showarticletitle{Improving pairwise learning for item recommendation
  from implicit feedback}. In \bibinfo{booktitle}{\emph{WSDM}}.
  \bibinfo{pages}{273--282}.
\newblock


\bibitem[\protect\citeauthoryear{Rendle, Freudenthaler, Gantner, and
  Schmidt-Thieme}{Rendle et~al\mbox{.}}{2009}]%
        {rendle2012bpr}
\bibfield{author}{\bibinfo{person}{Steffen Rendle}, \bibinfo{person}{Christoph
  Freudenthaler}, \bibinfo{person}{Zeno Gantner}, {and} \bibinfo{person}{Lars
  Schmidt-Thieme}.} \bibinfo{year}{2009}\natexlab{}.
\newblock \showarticletitle{BPR: Bayesian personalized ranking from implicit
  feedback}. In \bibinfo{booktitle}{\emph{UAI}}. \bibinfo{pages}{452--461}.
\newblock


\bibitem[\protect\citeauthoryear{Shi, Ma, Zhang, Zhang, Yu, Shan, Liu, and
  Ma}{Shi et~al\mbox{.}}{2020}]%
        {shi2020beyond}
\bibfield{author}{\bibinfo{person}{Shaoyun Shi}, \bibinfo{person}{Weizhi Ma},
  \bibinfo{person}{Min Zhang}, \bibinfo{person}{Yongfeng Zhang},
  \bibinfo{person}{Xinxing Yu}, \bibinfo{person}{Houzhi Shan},
  \bibinfo{person}{Yiqun Liu}, {and} \bibinfo{person}{Shaoping Ma}.}
  \bibinfo{year}{2020}\natexlab{}.
\newblock \showarticletitle{Beyond User Embedding Matrix: Learning to Hash for
  Modeling Large-Scale Users in Recommendation}. In
  \bibinfo{booktitle}{\emph{SIGIR}}. \bibinfo{pages}{319--328}.
\newblock


\bibitem[\protect\citeauthoryear{Shi, Karatzoglou, Baltrunas, Larson, Hanjalic,
  and Oliver}{Shi et~al\mbox{.}}{2012a}]%
        {shi2012tfmap}
\bibfield{author}{\bibinfo{person}{Yue Shi}, \bibinfo{person}{Alexandros
  Karatzoglou}, \bibinfo{person}{Linas Baltrunas}, \bibinfo{person}{Martha
  Larson}, \bibinfo{person}{Alan Hanjalic}, {and} \bibinfo{person}{Nuria
  Oliver}.} \bibinfo{year}{2012}\natexlab{a}.
\newblock \showarticletitle{Tfmap: optimizing map for top-n context-aware
  recommendation}. In \bibinfo{booktitle}{\emph{SIGIR}}.
  \bibinfo{pages}{155--164}.
\newblock


\bibitem[\protect\citeauthoryear{Shi, Karatzoglou, Baltrunas, Larson, Oliver,
  and Hanjalic}{Shi et~al\mbox{.}}{2012b}]%
        {shi2012climf}
\bibfield{author}{\bibinfo{person}{Yue Shi}, \bibinfo{person}{Alexandros
  Karatzoglou}, \bibinfo{person}{Linas Baltrunas}, \bibinfo{person}{Martha
  Larson}, \bibinfo{person}{Nuria Oliver}, {and} \bibinfo{person}{Alan
  Hanjalic}.} \bibinfo{year}{2012}\natexlab{b}.
\newblock \showarticletitle{CLiMF: learning to maximize reciprocal rank with
  collaborative less-is-more filtering}. In \bibinfo{booktitle}{\emph{RecSys}}.
  \bibinfo{pages}{139--146}.
\newblock


\bibitem[\protect\citeauthoryear{Shi, Larson, and Hanjalic}{Shi
  et~al\mbox{.}}{2010}]%
        {shi2010list}
\bibfield{author}{\bibinfo{person}{Yue Shi}, \bibinfo{person}{Martha Larson},
  {and} \bibinfo{person}{Alan Hanjalic}.} \bibinfo{year}{2010}\natexlab{}.
\newblock \showarticletitle{List-wise learning to rank with matrix
  factorization for collaborative filtering}. In
  \bibinfo{booktitle}{\emph{RecSys}}. \bibinfo{pages}{269--272}.
\newblock


\bibitem[\protect\citeauthoryear{Steck}{Steck}{2011}]%
        {steck2011item}
\bibfield{author}{\bibinfo{person}{Harald Steck}.}
  \bibinfo{year}{2011}\natexlab{}.
\newblock \showarticletitle{Item popularity and recommendation accuracy}. In
  \bibinfo{booktitle}{\emph{RecSys}}. \bibinfo{pages}{125--132}.
\newblock


\bibitem[\protect\citeauthoryear{Sun, Wu, and Wang}{Sun et~al\mbox{.}}{2018}]%
        {sun2018attentive}
\bibfield{author}{\bibinfo{person}{Peijie Sun}, \bibinfo{person}{Le Wu}, {and}
  \bibinfo{person}{Meng Wang}.} \bibinfo{year}{2018}\natexlab{}.
\newblock \showarticletitle{Attentive recurrent social recommendation}. In
  \bibinfo{booktitle}{\emph{SIGIR}}. \bibinfo{pages}{185--194}.
\newblock


\bibitem[\protect\citeauthoryear{Wang, Zhu, Zhu, Qin, and Xiong}{Wang
  et~al\mbox{.}}{2020}]%
        {wang2020setrank}
\bibfield{author}{\bibinfo{person}{Chao Wang}, \bibinfo{person}{Hengshu Zhu},
  \bibinfo{person}{Chen Zhu}, \bibinfo{person}{Chuan Qin}, {and}
  \bibinfo{person}{Hui Xiong}.} \bibinfo{year}{2020}\natexlab{}.
\newblock \showarticletitle{SetRank: A Setwise Bayesian Approach for
  Collaborative Ranking from Implicit Feedback.}. In
  \bibinfo{booktitle}{\emph{AAAI}}. \bibinfo{pages}{6127--6136}.
\newblock


\bibitem[\protect\citeauthoryear{Wang, Kim, McCord-Snook, Wu, and Wang}{Wang
  et~al\mbox{.}}{2019c}]%
        {wang2019variance}
\bibfield{author}{\bibinfo{person}{Huazheng Wang}, \bibinfo{person}{Sonwoo
  Kim}, \bibinfo{person}{Eric McCord-Snook}, \bibinfo{person}{Qingyun Wu},
  {and} \bibinfo{person}{Hongning Wang}.} \bibinfo{year}{2019}\natexlab{c}.
\newblock \showarticletitle{Variance reduction in gradient exploration for
  online learning to rank}. In \bibinfo{booktitle}{\emph{SIGIR}}.
  \bibinfo{pages}{835--844}.
\newblock


\bibitem[\protect\citeauthoryear{Wang, Chen, Zhu, Shen, and Zhang}{Wang
  et~al\mbox{.}}{2019a}]%
        {wang2019unified}
\bibfield{author}{\bibinfo{person}{Pengfei Wang}, \bibinfo{person}{Hanxiong
  Chen}, \bibinfo{person}{Yadong Zhu}, \bibinfo{person}{Huawei Shen}, {and}
  \bibinfo{person}{Yongfeng Zhang}.} \bibinfo{year}{2019}\natexlab{a}.
\newblock \showarticletitle{Unified collaborative filtering over graph
  embeddings}. In \bibinfo{booktitle}{\emph{SIGIR}}. \bibinfo{pages}{155--164}.
\newblock


\bibitem[\protect\citeauthoryear{Wang, Huang, Liu, Ma, Chen, and
  Veijalainen}{Wang et~al\mbox{.}}{2016}]%
        {wang2016ranking}
\bibfield{author}{\bibinfo{person}{Shuaiqiang Wang}, \bibinfo{person}{Shanshan
  Huang}, \bibinfo{person}{Tie-Yan Liu}, \bibinfo{person}{Jun Ma},
  \bibinfo{person}{Zhumin Chen}, {and} \bibinfo{person}{Jari Veijalainen}.}
  \bibinfo{year}{2016}\natexlab{}.
\newblock \showarticletitle{Ranking-oriented collaborative filtering: A
  listwise approach}.
\newblock \bibinfo{journal}{\emph{TOIS}} \bibinfo{volume}{35},
  \bibinfo{number}{2} (\bibinfo{year}{2016}), \bibinfo{pages}{1--28}.
\newblock


\bibitem[\protect\citeauthoryear{Wang, He, Wang, Feng, and Chua}{Wang
  et~al\mbox{.}}{2019b}]%
        {wang2019neural}
\bibfield{author}{\bibinfo{person}{Xiang Wang}, \bibinfo{person}{Xiangnan He},
  \bibinfo{person}{Meng Wang}, \bibinfo{person}{Fuli Feng}, {and}
  \bibinfo{person}{Tat-Seng Chua}.} \bibinfo{year}{2019}\natexlab{b}.
\newblock \showarticletitle{Neural graph collaborative filtering}. In
  \bibinfo{booktitle}{\emph{SIGIR}}. \bibinfo{pages}{165--174}.
\newblock


\bibitem[\protect\citeauthoryear{Wu, Volkovs, Soon, Sanner, and Rai}{Wu
  et~al\mbox{.}}{2019}]%
        {wu2019noise}
\bibfield{author}{\bibinfo{person}{Ga Wu}, \bibinfo{person}{Maksims Volkovs},
  \bibinfo{person}{Chee~Loong Soon}, \bibinfo{person}{Scott Sanner}, {and}
  \bibinfo{person}{Himanshu Rai}.} \bibinfo{year}{2019}\natexlab{}.
\newblock \showarticletitle{Noise Contrastive Estimation for One-Class
  Collaborative Filtering}. In \bibinfo{booktitle}{\emph{SIGIR}}.
  \bibinfo{pages}{135--144}.
\newblock


\bibitem[\protect\citeauthoryear{Wu, He, Wang, Zhang, and Wang}{Wu
  et~al\mbox{.}}{2021}]%
        {wu2021survey}
\bibfield{author}{\bibinfo{person}{Le Wu}, \bibinfo{person}{Xiangnan He},
  \bibinfo{person}{Xiang Wang}, \bibinfo{person}{Kun Zhang}, {and}
  \bibinfo{person}{Meng Wang}.} \bibinfo{year}{2021}\natexlab{}.
\newblock \showarticletitle{A Survey on Neural Recommendation: From
  Collaborative Filtering to Content and Context Enriched Recommendation}.
\newblock \bibinfo{journal}{\emph{arXiv preprint arXiv:2104.13030}}
  (\bibinfo{year}{2021}).
\newblock


\bibitem[\protect\citeauthoryear{Wu, Hsieh, and Sharpnack}{Wu
  et~al\mbox{.}}{2018}]%
        {wu2018sql}
\bibfield{author}{\bibinfo{person}{Liwei Wu}, \bibinfo{person}{Cho-Jui Hsieh},
  {and} \bibinfo{person}{James Sharpnack}.} \bibinfo{year}{2018}\natexlab{}.
\newblock \showarticletitle{SQL-Rank: A Listwise Approach to Collaborative
  Ranking}. In \bibinfo{booktitle}{\emph{ICML}}. \bibinfo{pages}{5315--5324}.
\newblock


\bibitem[\protect\citeauthoryear{Wu, Yang, Zhang, Hong, Fu, and Wang}{Wu
  et~al\mbox{.}}{2020}]%
        {wu2020joint}
\bibfield{author}{\bibinfo{person}{Le Wu}, \bibinfo{person}{Yonghui Yang},
  \bibinfo{person}{Kun Zhang}, \bibinfo{person}{Richang Hong},
  \bibinfo{person}{Yanjie Fu}, {and} \bibinfo{person}{Meng Wang}.}
  \bibinfo{year}{2020}\natexlab{}.
\newblock \showarticletitle{Joint item recommendation and attribute inference:
  An adaptive graph convolutional network approach}. In
  \bibinfo{booktitle}{\emph{SIGIR}}. \bibinfo{pages}{679--688}.
\newblock


\bibitem[\protect\citeauthoryear{Xia, Liu, Wang, Zhang, and Li}{Xia
  et~al\mbox{.}}{2008}]%
        {xia2008listwise}
\bibfield{author}{\bibinfo{person}{Fen Xia}, \bibinfo{person}{Tie-Yan Liu},
  \bibinfo{person}{Jue Wang}, \bibinfo{person}{Wensheng Zhang}, {and}
  \bibinfo{person}{Hang Li}.} \bibinfo{year}{2008}\natexlab{}.
\newblock \showarticletitle{Listwise approach to learning to rank: theory and
  algorithm}. In \bibinfo{booktitle}{\emph{ICML}}. \bibinfo{pages}{1192--1199}.
\newblock


\bibitem[\protect\citeauthoryear{Xue, Dai, Zhang, Huang, and Chen}{Xue
  et~al\mbox{.}}{2017}]%
        {xue2017deep}
\bibfield{author}{\bibinfo{person}{Hong-Jian Xue}, \bibinfo{person}{Xinyu Dai},
  \bibinfo{person}{Jianbing Zhang}, \bibinfo{person}{Shujian Huang}, {and}
  \bibinfo{person}{Jiajun Chen}.} \bibinfo{year}{2017}\natexlab{}.
\newblock \showarticletitle{Deep Matrix Factorization Models for Recommender
  Systems.}. In \bibinfo{booktitle}{\emph{IJCAI}}. \bibinfo{pages}{3203--3209}.
\newblock


\bibitem[\protect\citeauthoryear{Yu, Bilenko, and Lin}{Yu
  et~al\mbox{.}}{2017}]%
        {yu2017selection}
\bibfield{author}{\bibinfo{person}{Hsiang-Fu Yu}, \bibinfo{person}{Mikhail
  Bilenko}, {and} \bibinfo{person}{Chih-Jen Lin}.}
  \bibinfo{year}{2017}\natexlab{}.
\newblock \showarticletitle{Selection of negative samples for one-class matrix
  factorization}. In \bibinfo{booktitle}{\emph{SIAM}}.
  \bibinfo{pages}{363--371}.
\newblock


\bibitem[\protect\citeauthoryear{Yu, Liu, Ye, Cheng, Chen, and Ma}{Yu
  et~al\mbox{.}}{2020}]%
        {yu2020collaborative}
\bibfield{author}{\bibinfo{person}{Runlong Yu}, \bibinfo{person}{Qi Liu},
  \bibinfo{person}{Yuyang Ye}, \bibinfo{person}{Mingyue Cheng},
  \bibinfo{person}{Enhong Chen}, {and} \bibinfo{person}{Jianhui Ma}.}
  \bibinfo{year}{2020}\natexlab{}.
\newblock \showarticletitle{Collaborative List-and-Pairwise Filtering from
  Implicit Feedback}.
\newblock \bibinfo{journal}{\emph{TKDE}} (\bibinfo{year}{2020}).
\newblock


\bibitem[\protect\citeauthoryear{Yu, Zhang, Ye, Wu, Wang, Liu, and Chen}{Yu
  et~al\mbox{.}}{2018}]%
        {yu2018multiple}
\bibfield{author}{\bibinfo{person}{Runlong Yu}, \bibinfo{person}{Yunzhou
  Zhang}, \bibinfo{person}{Yuyang Ye}, \bibinfo{person}{Le Wu},
  \bibinfo{person}{Chao Wang}, \bibinfo{person}{Qi Liu}, {and}
  \bibinfo{person}{Enhong Chen}.} \bibinfo{year}{2018}\natexlab{}.
\newblock \showarticletitle{Multiple pairwise ranking with implicit feedback}.
  In \bibinfo{booktitle}{\emph{CIKM}}. \bibinfo{pages}{1727--1730}.
\newblock


\bibitem[\protect\citeauthoryear{Yu and Qin}{Yu and Qin}{2020}]%
        {yu2020sampler}
\bibfield{author}{\bibinfo{person}{Wenhui Yu} {and} \bibinfo{person}{Zheng
  Qin}.} \bibinfo{year}{2020}\natexlab{}.
\newblock \showarticletitle{Sampler Design for Implicit Feedback Data by
  Noisy-label Robust Learning}. In \bibinfo{booktitle}{\emph{SIGIR}}.
  \bibinfo{pages}{861--870}.
\newblock


\bibitem[\protect\citeauthoryear{Zarzour, Al-Sharif, Al-Ayyoub, and
  Jararweh}{Zarzour et~al\mbox{.}}{2018}]%
        {zarzour2018new}
\bibfield{author}{\bibinfo{person}{Hafed Zarzour}, \bibinfo{person}{Ziad
  Al-Sharif}, \bibinfo{person}{Mahmoud Al-Ayyoub}, {and} \bibinfo{person}{Yaser
  Jararweh}.} \bibinfo{year}{2018}\natexlab{}.
\newblock \showarticletitle{A new collaborative filtering recommendation
  algorithm based on dimensionality reduction and clustering techniques}. In
  \bibinfo{booktitle}{\emph{ICICS}}. \bibinfo{pages}{102--106}.
\newblock


\bibitem[\protect\citeauthoryear{Zhang, Chen, Wang, and Yu}{Zhang
  et~al\mbox{.}}{2013}]%
        {zhang2013optimizing}
\bibfield{author}{\bibinfo{person}{Weinan Zhang}, \bibinfo{person}{Tianqi
  Chen}, \bibinfo{person}{Jun Wang}, {and} \bibinfo{person}{Yong Yu}.}
  \bibinfo{year}{2013}\natexlab{}.
\newblock \showarticletitle{Optimizing top-n collaborative filtering via
  dynamic negative item sampling}. In \bibinfo{booktitle}{\emph{SIGIR}}.
  \bibinfo{pages}{785--788}.
\newblock


\bibitem[\protect\citeauthoryear{Zhong, Pan, Xu, Yin, and Ming}{Zhong
  et~al\mbox{.}}{2014}]%
        {zhong2014adaptive}
\bibfield{author}{\bibinfo{person}{Hao Zhong}, \bibinfo{person}{Weike Pan},
  \bibinfo{person}{Congfu Xu}, \bibinfo{person}{Zhi Yin}, {and}
  \bibinfo{person}{Zhong Ming}.} \bibinfo{year}{2014}\natexlab{}.
\newblock \showarticletitle{Adaptive pairwise preference learning for
  collaborative recommendation with implicit feedbacks}. In
  \bibinfo{booktitle}{\emph{CIKM}}. \bibinfo{pages}{1999--2002}.
\newblock


\end{thebibliography}

% \appendix

\end{document}